\documentclass[12pt,english,a4paper]{article}
\usepackage{natbib}
\usepackage{rotating,placeins}
\usepackage{babel}
\usepackage{geometry}
\usepackage{fancyhdr}
\usepackage{lscape,epsfig,amssymb,amsmath}
\usepackage[]{subfigure}
\usepackage{rotating,placeins}
\usepackage{lscape}
\usepackage{setspace}
\newcommand{\Hs}{$H_\mathrm{s}$}
\newcommand{\Hr}{$H_{100}$}
\newcommand{\Hdre}{$H^\mathrm{DRE}_{100}$}
\newcommand{\Hexp}{$H^\mathrm{EXP}_{100}$}
\newcommand{\Hgp}{$H^\mathrm{GP}_{100}$}

\newcommand{\U}{$U_\mathrm{10m}$}
\newcommand{\Ur}{$U_{100}$}
\newcommand{\Udre}{$U^\mathrm{DRE}_{100}$}
\newcommand{\Uexp}{$U^\mathrm{EXP}_{100}$}
\newcommand{\Ugp}{$U^\mathrm{GP}_{100}$}

\newcommand{\Fig}[1]{Fig~\ref{fig:#1}}
\newcommand{\Eq}[1]{Eq~(\ref{eq:#1})}

\title{Wind and Wave Extremes over the World Oceans from Very Large Ensembles}
\author{{\O}yvind Breivik\footnote{Final version published in
\textit{Geophys Res Lett}, 
       doi:10.1002/2014GL060997}\thanks{Corresponding author. E-mail:
          \texttt{oyvind.breivik@ecmwf.int}. 
          ORCID Author ID: \texttt{0000-0002-2900-8458}}
  \thanks{ECMWF, Shinfield Park, Reading, RG2 9AX, UK.}
  \and Ole Johan Aarnes\thanks{Norwegian Meteorological Institute.}
  \and Saleh Abdalla\footnotemark[3]
  \and Jean-Raymond Bidlot\footnotemark[3]
  \and Peter A.E.M. Janssen\footnotemark[3]}

\begin{document}
\maketitle

\abstract{
Global return values of marine wind speed and significant wave height
are estimated from very large aggregates of archived
ensemble forecasts at +240-h lead time. Long lead time ensures that the
forecasts represent independent draws from the model climate. Compared with
ERA-Interim, a reanalysis, the ensemble yields higher return estimates for
both wind speed and significant wave height.  Confidence intervals are much
tighter due to the large size of the dataset.  The period (9 yrs) is short
enough to be considered stationary even with climate change.  Furthermore,
the ensemble is large enough for non-parametric 100-yr return estimates
to be made from order statistics. These direct return estimates compare
well with extreme value estimates outside areas with tropical cyclones.
Like any method employing modeled fields, it is sensitive to tail biases in
the numerical model, but we find that the biases are moderate outside areas
with tropical cyclones.}

%Intro
\section{The ECMWF Integrated Forecast System Ensemble}
\label{sec:intro}
Return values for wind and waves are fundamental to assessing the
risks associated with human activities at sea, but their computation
is complicated by the paucity of observational records.  A number of
reanalyses and hindcasts have appeared over the past two decades, both global
\citep{kalnay96,upp05,onogi07,rienecker11,dee11,saha13} and regional, e.g.
\citet{wang01,wang02,weisse07,bre09,rei11,wang12}. However, with few exceptions
\citep{compo11,hersbach13} the time series are much shorter than the 100-yr
return period typically sought and any extremes estimated from them come
with wide confidence intervals (see \citet{aar12} and \citet{bre13b}, the
latter hereafter B13).

The Integrated Forecast System (IFS) of the European Centre for Medium-Range
Weather Forecasts (ECMWF) has been producing daily ensemble forecasts since
1992 \citep{buizza07} and has been coupled to the ECMWF version of the WAM wave
model since 1998 \citep{jan04}. 51-member ensemble forecasts have been issued
twice daily (00 and 12 UTC) since March 2003. Even though the forecast skill
has steadily been improving over the years \citep{magnusson13}, ensemble
members at +240~h lead time (ENS240 hereafter) still tend to be weakly
correlated with one another, and storm events are virtually uncorrelated,
as demonstrated by B13.  Although the goal of forecasting will always be to
raise the skill, perhaps somewhat paradoxically such weak correlations are a
necessity when using ensemble forecasts for extreme value estimation because
the entries must be independent \citep{col01}. That the tail of the ensemble
can be considered independent was demonstrated by B13 where memberwise return
estimates were shown to give very similar results to return estimates from
the entire ensemble. This also showed that the correlation between
consecutive +240-h forecasts is negligible.

%Comparison_with_upper_percentiles_of_observations_and_ENVISAT_and_ERA-I
Although the forecast skill must be low for the ensemble to be used for return
value estimation, the empirical forecast distribution must nevertheless closely
match the observed distribution all the way to the tail to yield realistic
return values. Hence the tail bias, i.e., the difference between observed
and modeled upper percentiles, should be as low as possible. We find very
good agreement with the upper percentiles of buoy wave and wind observations
and ENVISAT altimeter wind speed measurements (see Figs
\ref{fig:P99}-\ref{fig:envisat}).
It is also of interest to compare the tail behavior of ENS240 with ERA-Interim
(ERA-I) as it is one of the most widely used reanalyses, but also because it
is a version of IFS (Cycle 31r2) operational from December 2006 until June
2007 \citep{dee11}. It thus serves as a benchmark against which to compare
the ENS240 climatology. We find that ENS240 exhibits smaller biases than
ERA-I compared with altimeter wind speed measurements, with ERA-I typically
being biased low by more than $-1.5$ m s$^{-1}$ in certain regions and about
$-0.5$ m s$^{-1}$ globally.  This is in contrast to ENS240 which is virtually
unbiased when averaged globally and rarely deviates more than $\pm 0.5$
m~s$^{-1}$ locally (see \Fig{envisat}b).

\section{Estimating Return Values from Forecast Ensembles}
\label{sec:dre}
So far we have two conditions that must be met for the ensemble to be useful
for estimating probabilities of non-exceedance; \textbf{(i)} that the forecasts
are uncorrelated, and \textbf{(ii)} that the empirical distribution function
closely matches the observations all the way to the tail. To these we add
that \textbf{(iii)} the model climatology must be statistically stationary.
Since the archive spans numerous model cycles, this is not trivially true, but
B13 found no spurious trends in mean and variance over the period compared to
a reforecast ensemble from a fixed model cycle \citep{hagedorn12}.  One final
condition must be met to go from a probability of non-exceedance to a return
period: \textbf{(iv)} that individual forecasts are representative of a
synoptic (6~h) time interval. This allows us to treat 330,000 forecasts as
the equivalent of 229 yrs of data (see Appendix 1).

%Discussion_of_the_use_of_EVDs_v_DRE
Theoretical extreme value distributions are parametric estimates to be
fitted to the modeled or observed maxima that are normally obtained from a
continuous time series with a fixed temporal resolution.  Annual maxima (AM)
belong to the family of blocked maxima \citep{col01} and should follow the
Generalized Extreme Value distribution. Threshold exceedances follow the
Generalized Pareto (GP) distribution \citep{col01}, where in the case of
correlated time series a peaks-over-threshold (POT) method must be applied
to ensure that the maxima are independent of each other (see B13). With the
initial distribution method (IDM), all data are used.  Usually the method is
employed when data are scarce, but sensitivity to the mean and variance of
the series \citep{lop00} as well as a lack of theoretical justification for
a choice of distribution \citep{wmo98} makes the method highly empirical and
somewhat difficult to use. The choice of data selection method (AM, threshold
exceedances or IDM) and the choice of location and shape parameters for
the extreme value distribution can have a profound impact on the estimates
\citep{col01,aar12}, especially with time series significantly shorter than
the return period in question.

Because the ENS240 dataset is larger than the return period sought,
it is possible to avoid extreme value analysis altogether and compute
a non-parametric direct return estimate (DRE) from the tail of the
empirical distribution function (see Appendix).  Figs~\ref{fig:U100}a and
\ref{fig:h100}a show maps of the 100-yr return values of \Udre~and \Hdre,
respectively.  Estimates of \Ur~and \Hr~based on GP and the special case of
the exponential distribution (EXP) closely resemble DRE in the extratropics
(see Fig~\ref{fig:drediff} and \Fig{dre_v_exp}), with \Ugp~estimates about
1.0 m~s$^{-1}$ below \Udre.  Similarly, \Hgp~estimates are found to be within
$\pm0.5$~m of \Hdre~ for most of the extratropics (\Fig{drediff}b).  However,
in the tropics the parametric estimates deviate significantly from DRE.
The differences match closely areas visited by tropical cyclones, see eg
\citet{oouchi06}. EXP will yield significantly lower estimates compared with
DRE while GP does the opposite.  This is illustrated by \Fig{41001} where
the confidence interval of \Uexp~fails to encompass the extremal values
for a location in the subtropics. Tropical cyclones are underestimated in
global numerical weather prediction models \citep{magnusson14}, thus the GP
estimates come closer to the real return values in the affected areas, albeit
still with large uncertainties and a large negative bias. It is
interesting to note in passing that the highest wind speed found
throughout the entire ensemble is close to 50 m~s$^{-1}$ while \Hs~comes
close to 25~m (see \Fig{max}).

%Comparison_with_return_values_based_on_observations_and_ERA-I
\section{Comparison With Traditional Return Estimates}
\label{sec:comparison}
\begin{figure}[h]
\begin{center}
(a)\includegraphics[scale=0.6]{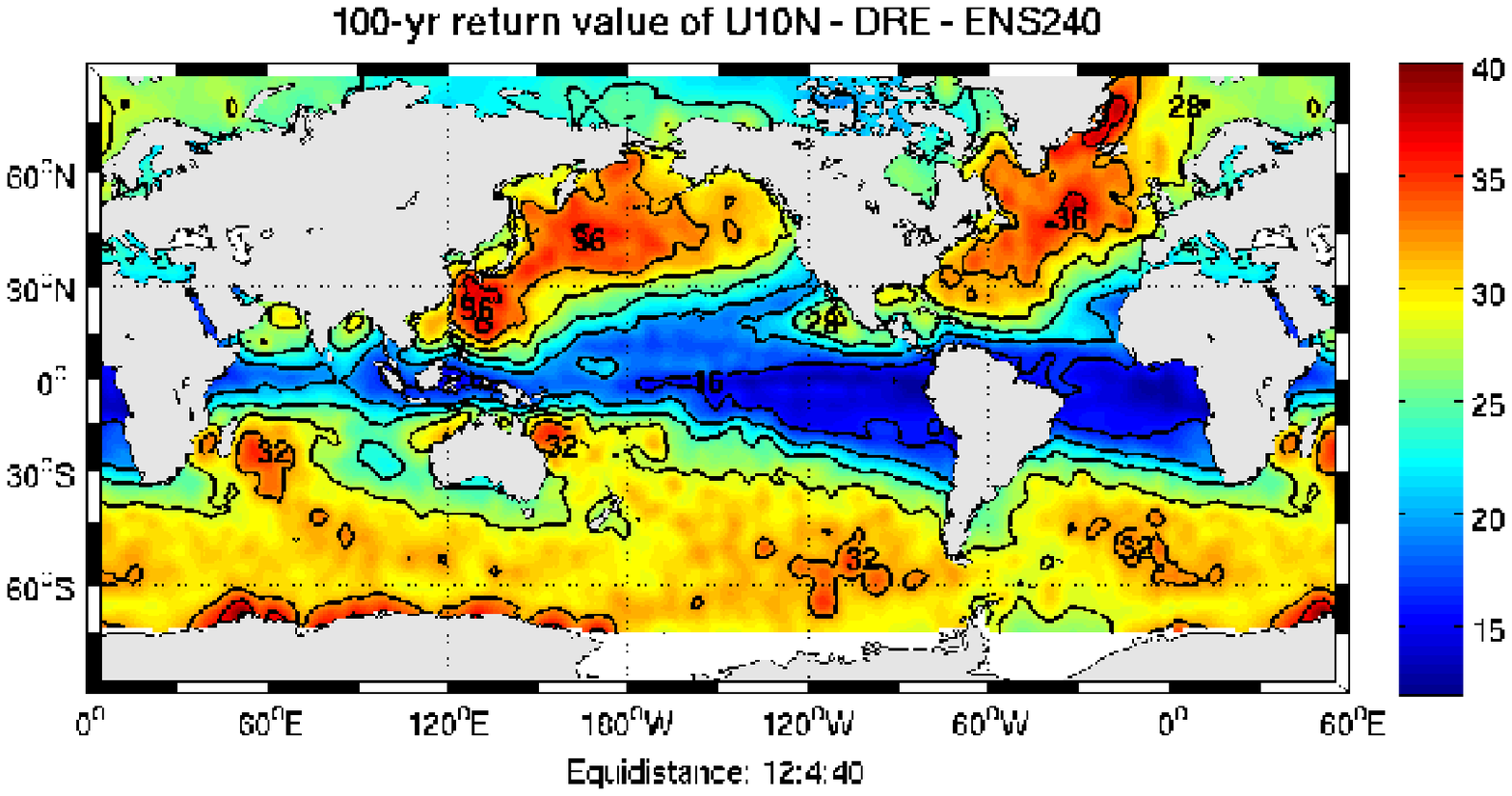}\\
(b)\includegraphics[scale=0.6]{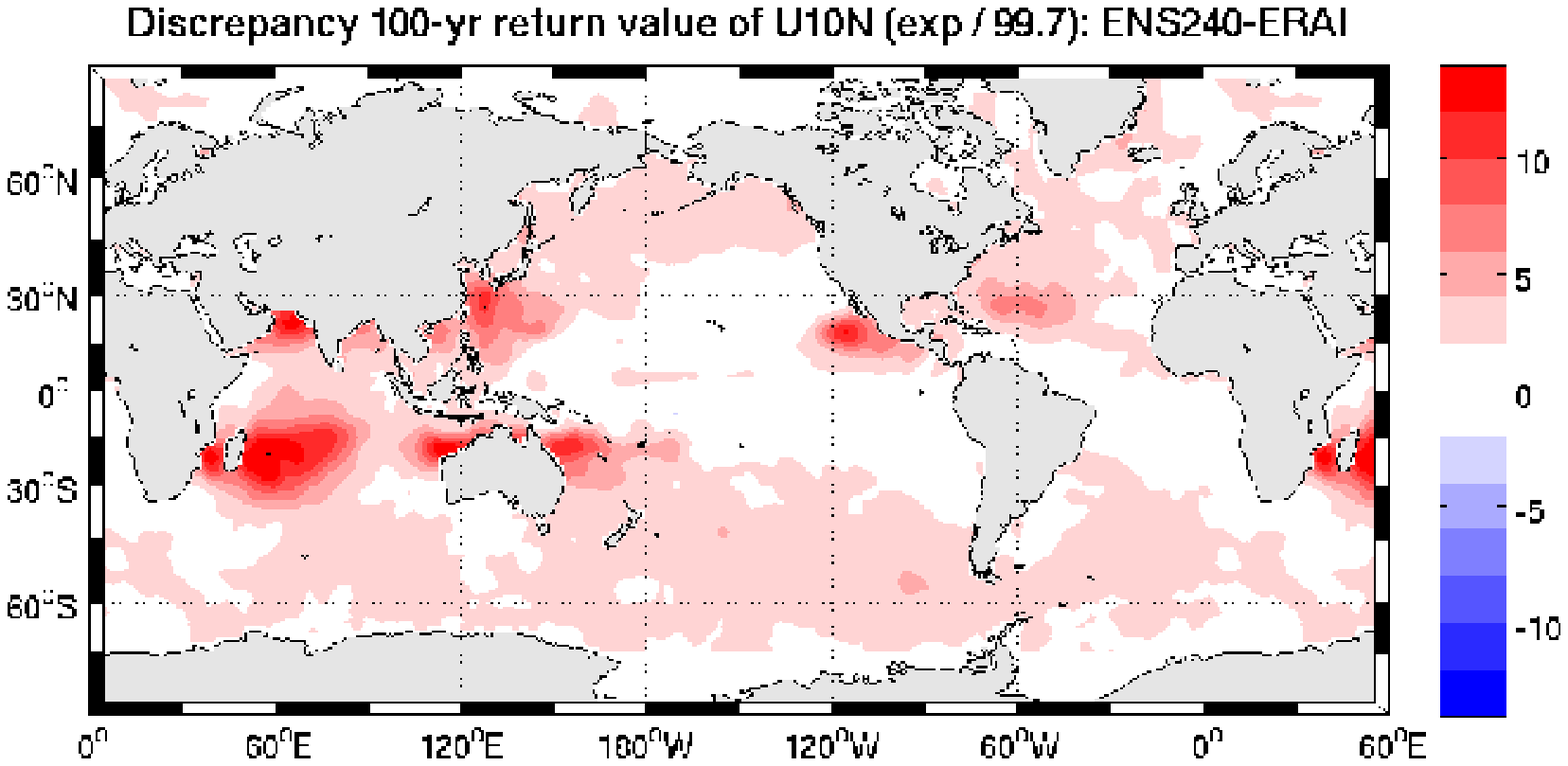}
\caption{
10-m wind speed 100-yr return values, \Udre~[m s$^{-1}$].  Panel a: ENS240
estimate direct return estimate. Panel b: Difference between ENS240 and ERA-I,
exponential distribution fit, threshold 99.7 percentile. 
} % end caption
\label{fig:U100} 
\end{center} 
\end{figure}

\begin{figure}[h]
\begin{center}
(a)\includegraphics[scale=0.7]{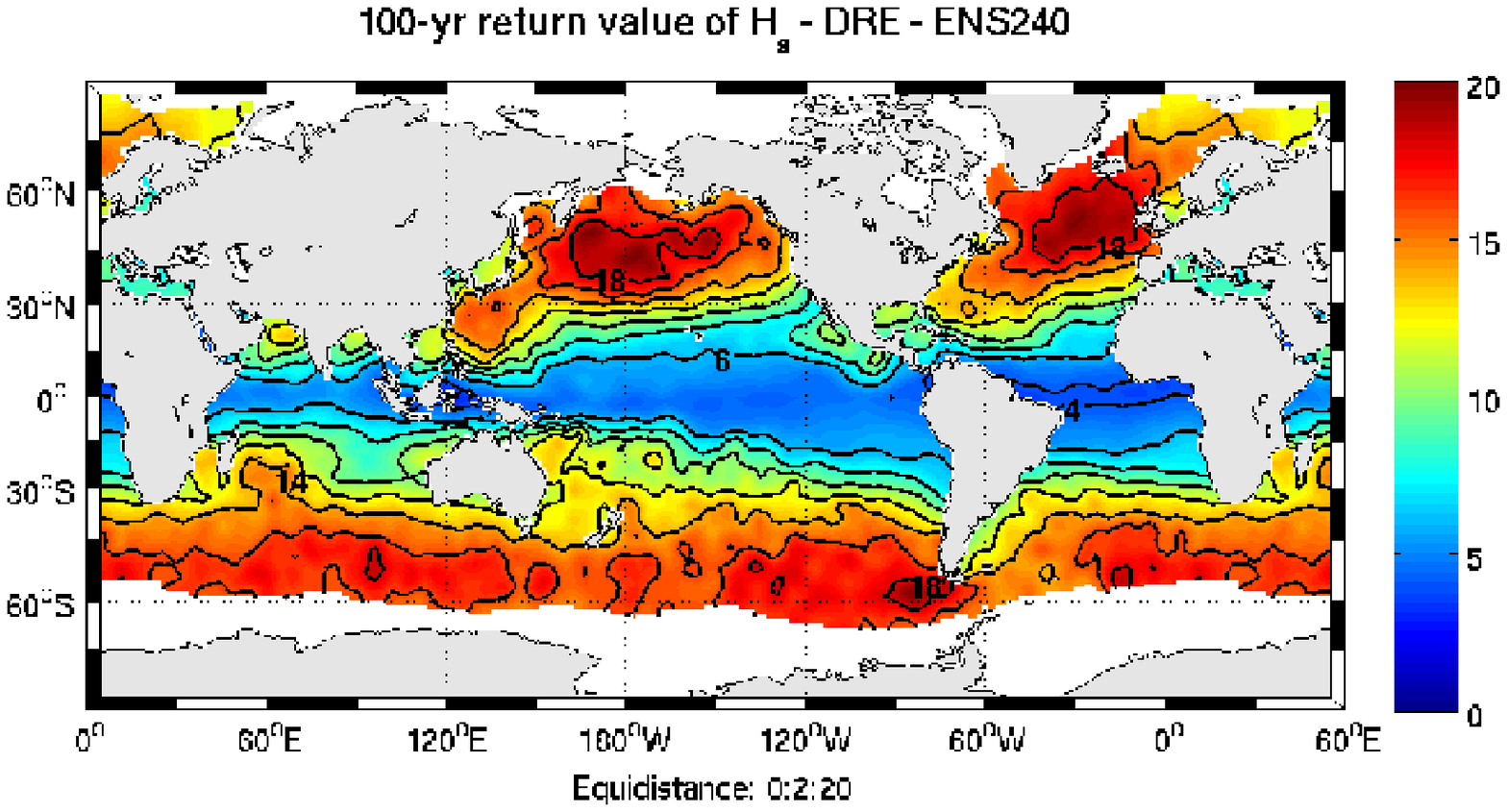}\\
(b)\includegraphics[scale=0.7]{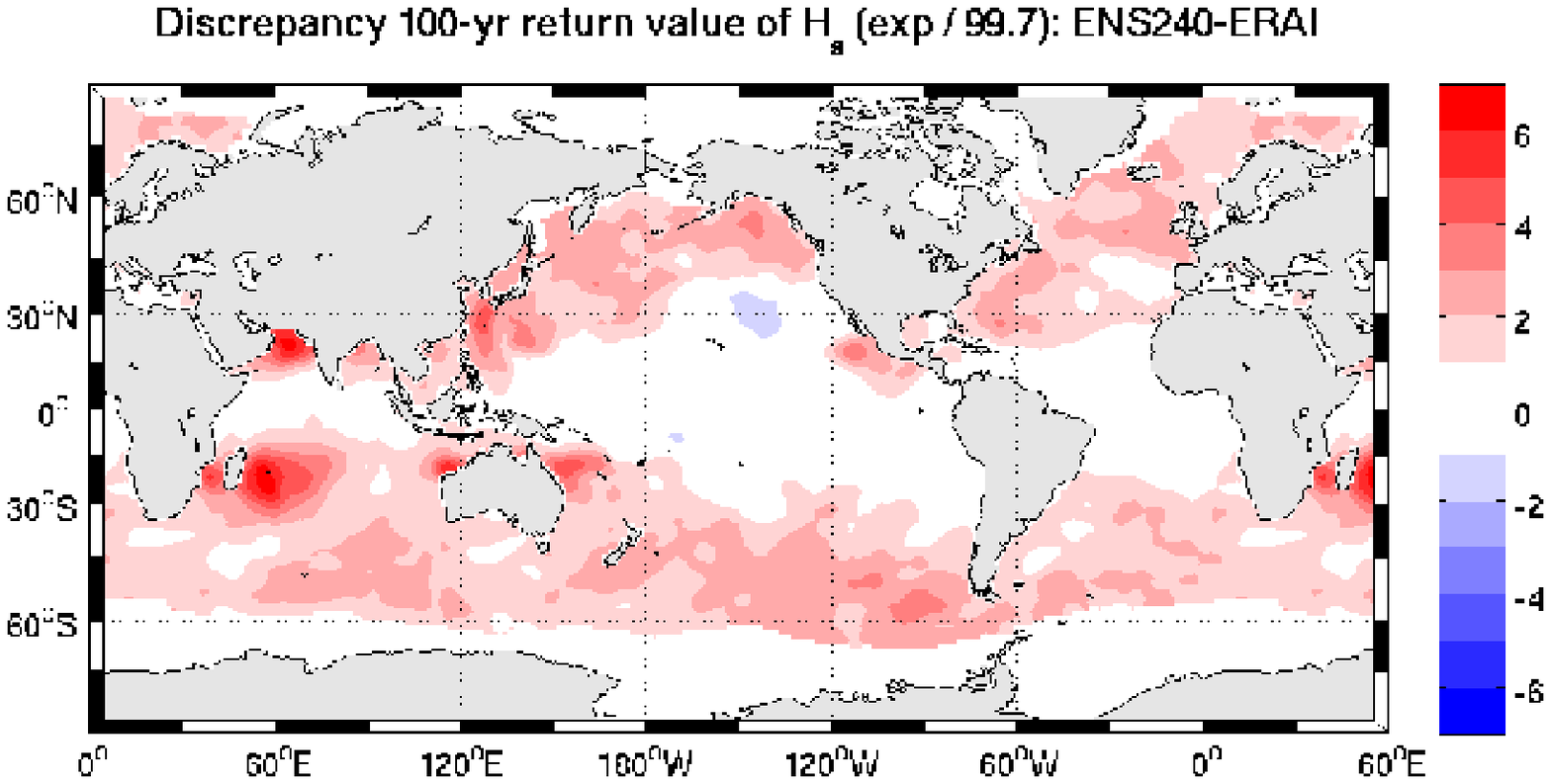}\\
\caption{
Significant wave height 100-yr return values, \Hr~[m]. Panel a: ENS240 direct
return estimate.  Panel b: Difference between ENS240 and ERA-I, exponential
distribution fit, threshold 99.7 percentile. Grid points that are ice-covered
more than 20\% of the time are censored.
} % end caption
\label{fig:h100} 
\end{center} 
\end{figure}

\begin{figure}[h]
\begin{center}
(a)\includegraphics[scale=0.7]{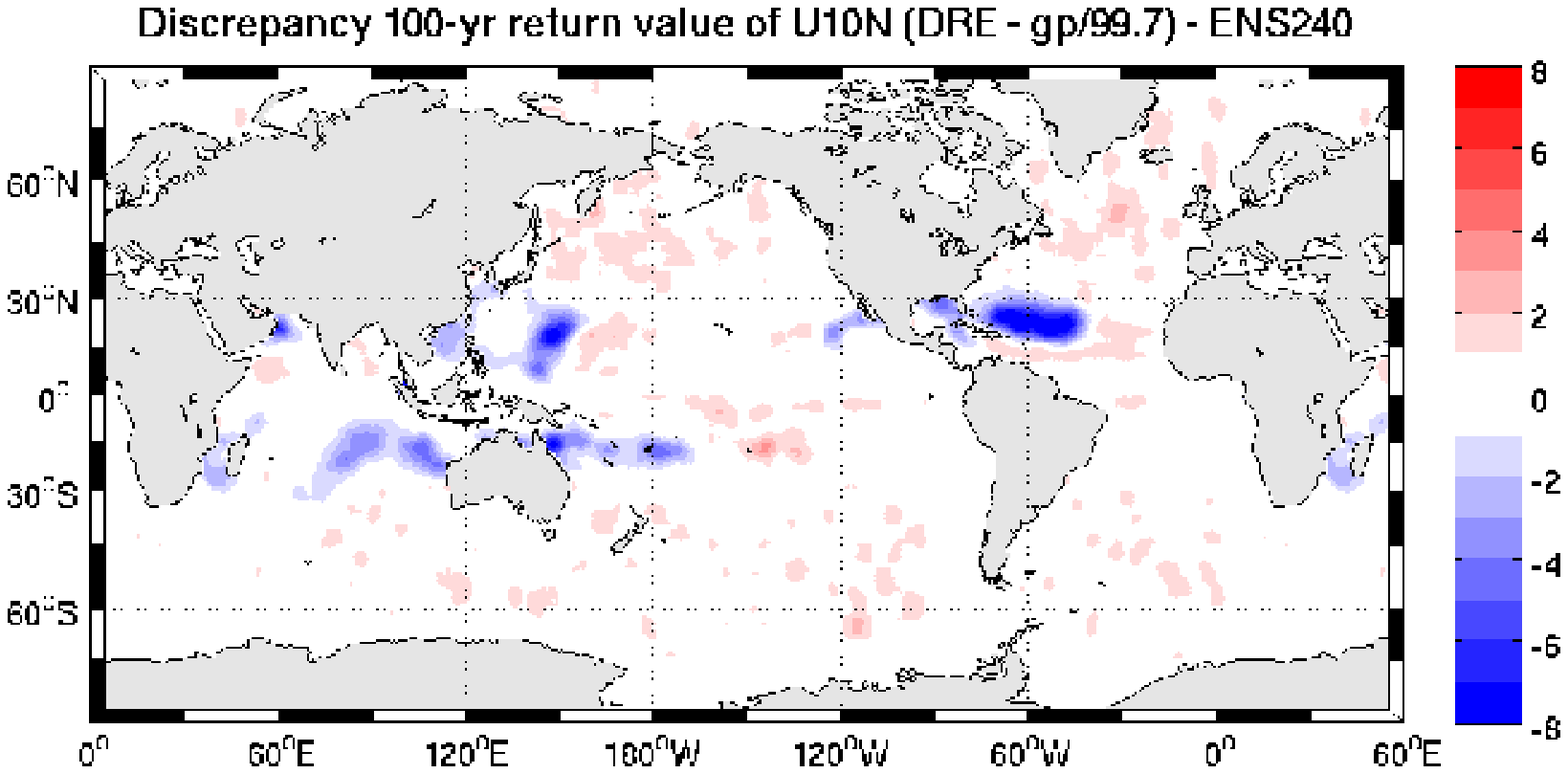}\\
(b)\includegraphics[scale=0.7]{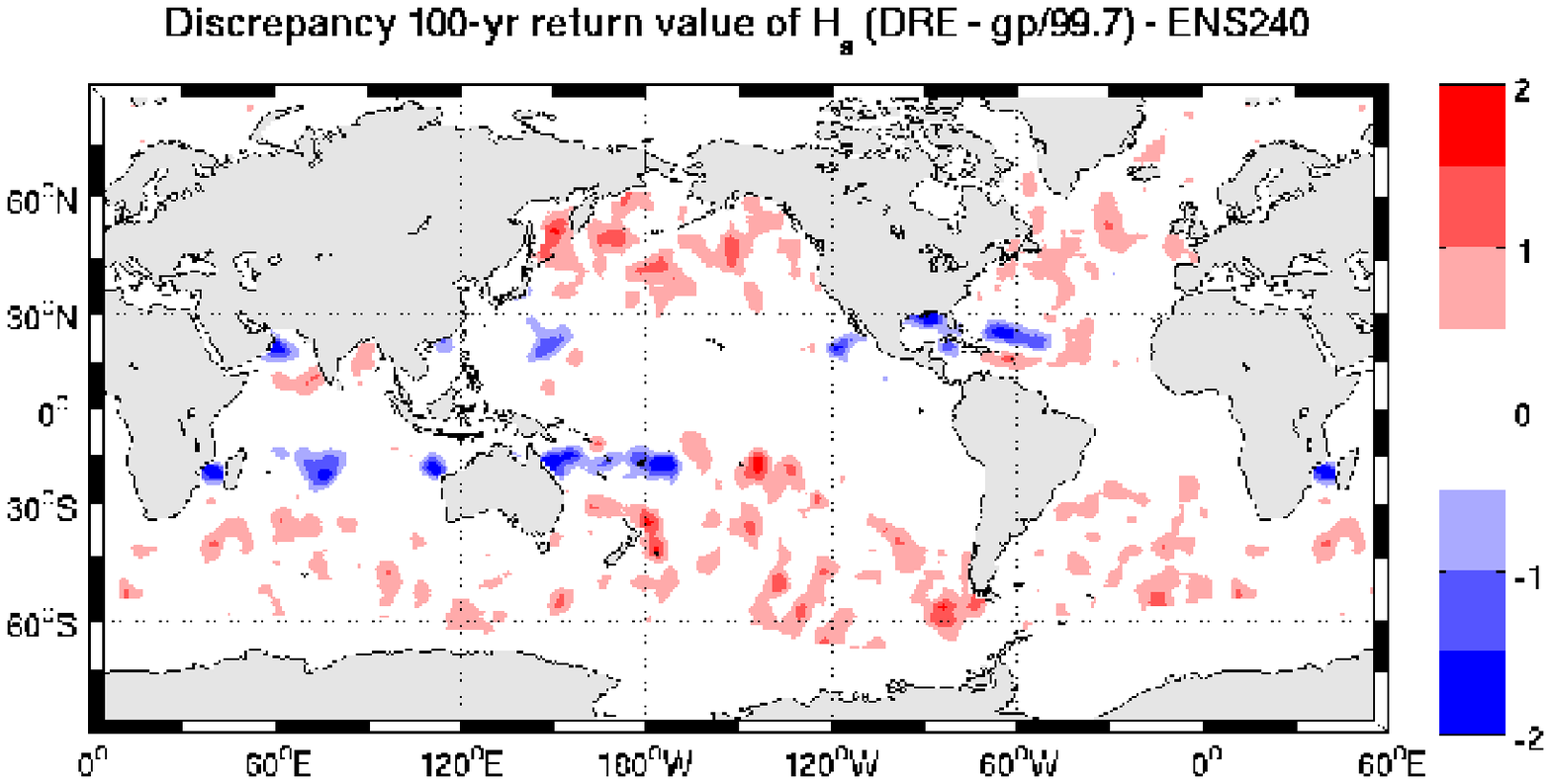}
\caption{
Non-parametric direct return estimates v parametric GP estimates.
Panel a: The difference between the \Udre~direct return estimate and the
\Ugp~GP estimate for ENS240 [m s$^{-1}$]. The differences are greatest in the
regions prone to tropical cyclones where GP has a positive shape parameter.
Panel b: Same as Panel a but for \Hr~[m]. Note that grid points that are
ice-covered more than 20\% of the time are censored for \Hs.
} % end caption
\label{fig:drediff}
\end{center} 
\end{figure}

\begin{figure}[h]
\begin{center}
\includegraphics[scale=0.5]{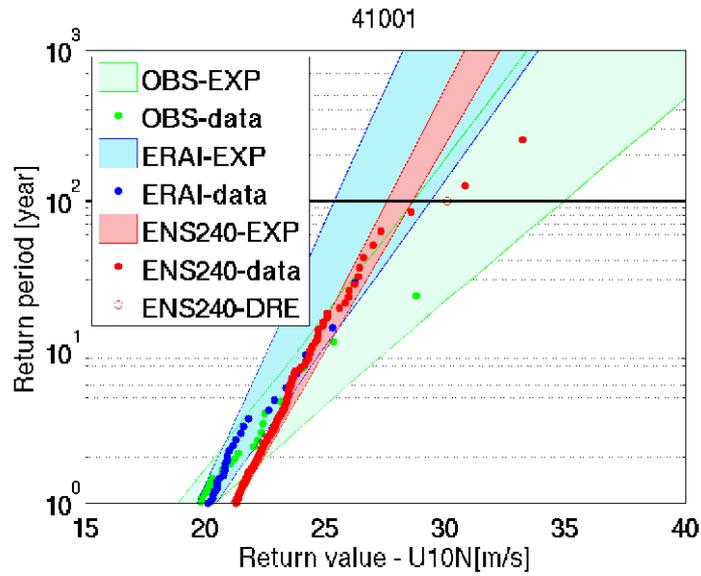}
\caption{
A comparison of the exponential distribution confidence intervals
of \Ur~for buoy 41001, located east of Florida at 34.7$^\circ$N,
072.7$^\circ$W. Bootstrapped (500 resamples) confidence intervals are shown
for \textit{in situ} measurements from 1980 to 2012 with some gaps (green),
ERA-I (blue) and ENS240 (red). The direct return estimate is shown as a red
open circle. The horizontal black line indicates the 100-yr return level.
As
expected the confidence intervals decrease with magnitude of the datasets,
but fail to include the highest data points. A GP distribution with a shape
parameter $\xi$ different from zero would fit the data better.  This is a
feature particularly prominent in areas where the tail of the wind speed
distribution is influenced by tropical cyclones.
} % end caption
\label{fig:41001} 
\end{center} 
\end{figure}

Because ERA-I and the \textit{in situ} observational time series are
substantially shorter than the return period sought, we compare with ENS240
in terms of parametric return value estimates.  ERA-I EXP estimates tend
to exhibit a more stable behavior than GP estimates, and will be used in
the following (see also Supplementary Material, Appendix \ref{sec:supp}).  In the extratropics the
ENS240 \Ur~ estimates are 2-4 m~s$^{-1}$ higher than ERA-I (\Fig{U100}b)
while \Hr~ estimates are 1-3 m higher (\Fig{h100}b).  The situation is
quite different in areas with tropical cyclones. Here we find much larger
differences (more than 14 m~s$^{-1}$ for \Ur, see \Fig{U100}b, and more
than 6 m for \Hr, see \Fig{h100}b).  However, it is clear that the coarse
resolution of the atmospheric model still renders tropical cyclones much
too weak \citep{magnusson14} compared to the maximum sustained wind speed
observed under such weather systems \citep{elsner08}.  This means that our
wind speed estimates in the subtropics and the tropics should be considered
lower bounds. It is worth remembering at this point, though, that the winds
reported here must be considered averages over much longer periods (6 h)
than the 10 minutes typically investigated in relation to tropical cyclones
\citep{bidlot02} (see Supplementary Material, Appendix \ref{sec:supp}).

\begin{figure}[h]
\begin{center}
(a)\includegraphics[scale=0.7]{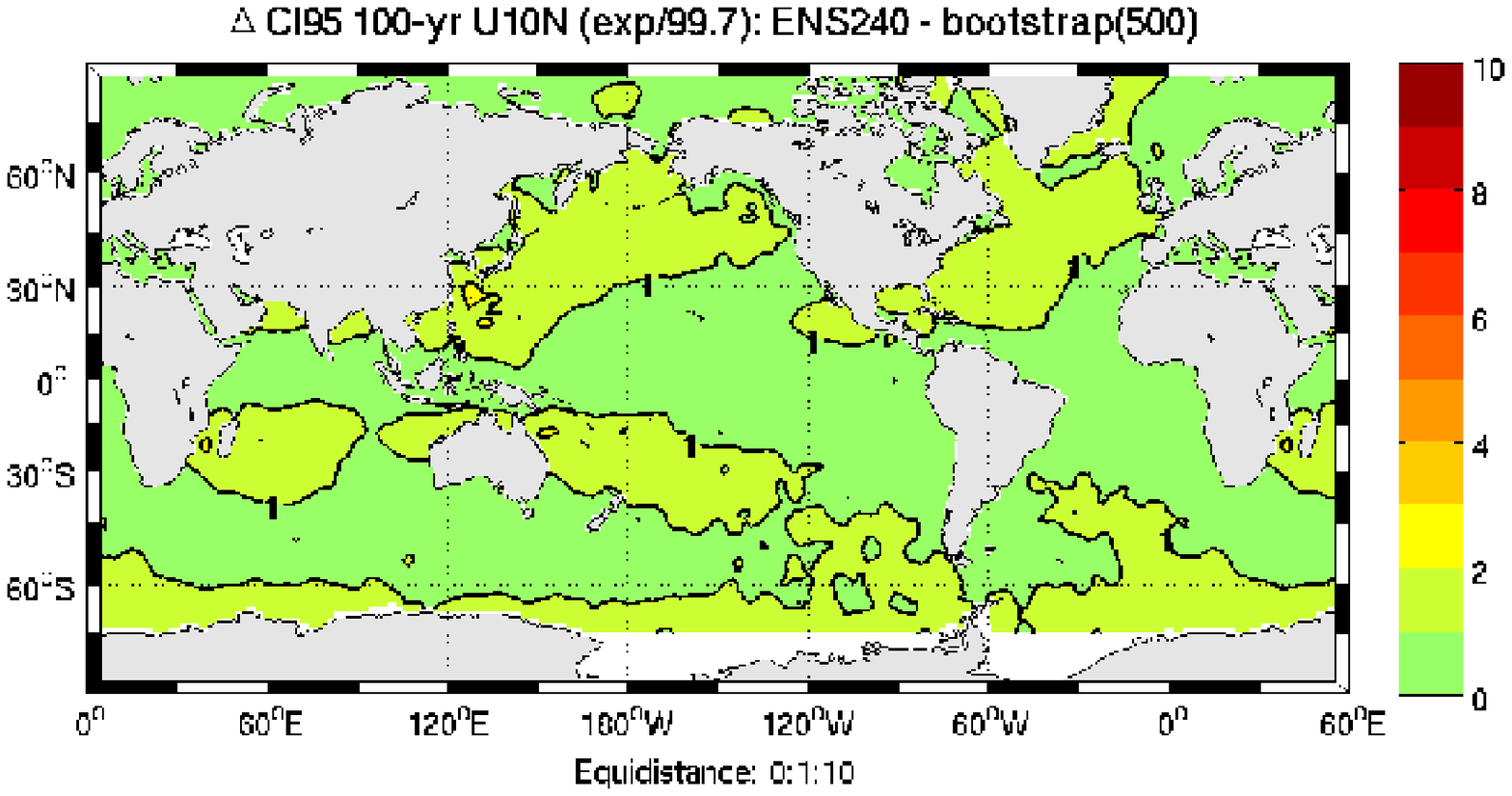}\\
(b)\includegraphics[scale=0.7]{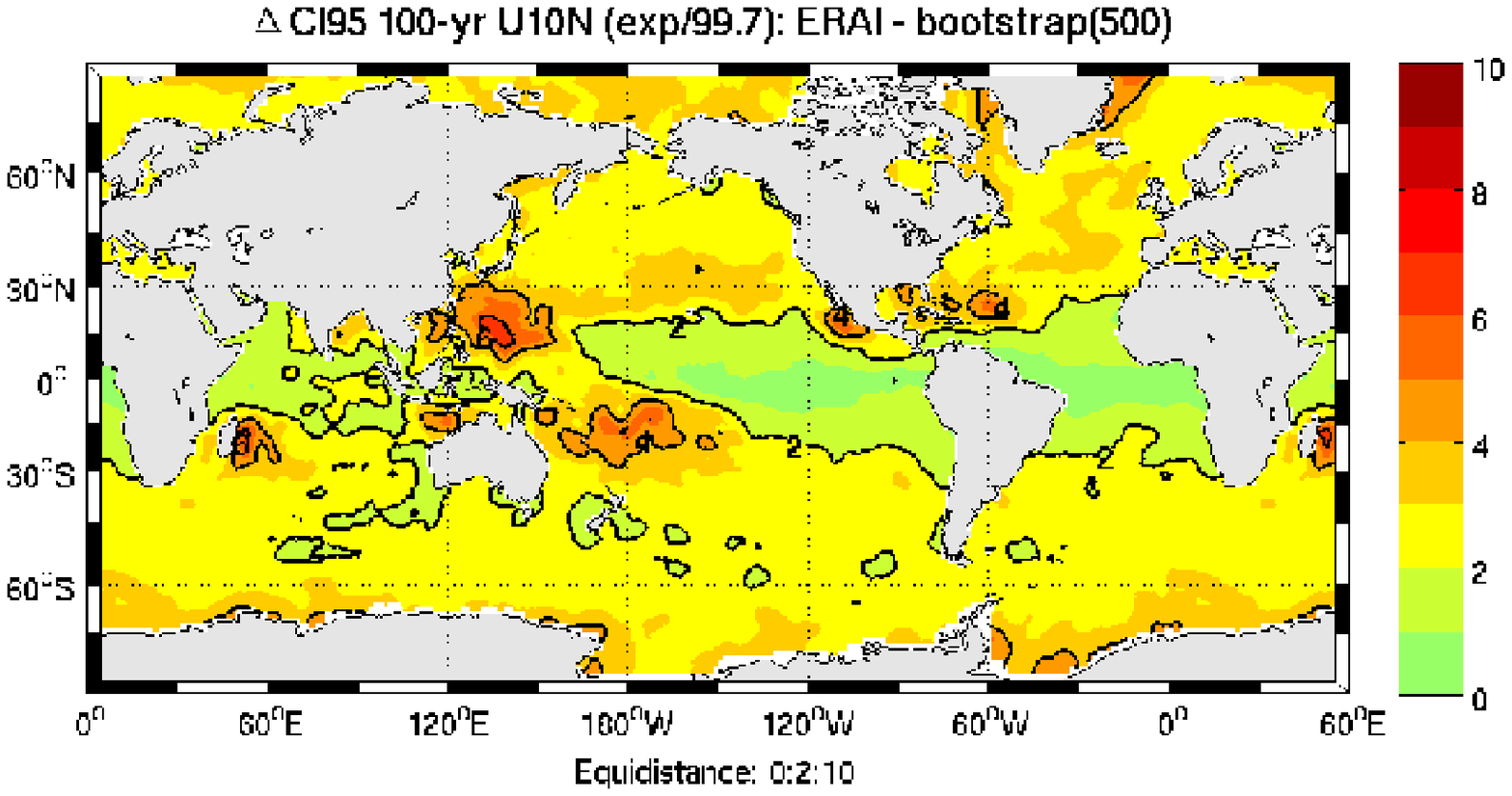}\\
\caption{
ENS240 v ERA-I width of \Uexp~ confidence intervals.  Estimates are fitted to
the EXP distribution, confidence intervals are computed using bootstrapping
of 500 resamples.  ENS240 has confidence intervals about one third the width
of ERA-I due to the much larger dataset.
Panel a: Width of the ENS240 \Uexp~ 95\% confidence interval. All data exceeding
$P_{99.7}$ were used.
Panel b: Same a panel a but for ERA-I. A peaks-over-threshold method was
used to select data exceeding a threshold of $P_{99.7}$.}
\label{fig:uexp_ci95_v_erai} 
\end{center} 
\end{figure}

ENS240 is assumed to represent the equivalent of approximately 229 yrs.
The size of the dataset reduces the confidence intervals to less than
one third the width of ERA-I confidence intervals (34-yr dataset), see
Figs~\ref{fig:uexp_ci95_v_erai} and \ref{fig:hexp_ci95_v_erai}.  In general,
ENS240 \Uexp~and \Hexp~compare reasonably well with POT estimates (see
Appendix) of \textit{in situ} observations (less than 10\% difference except in areas
with tropical cyclones, see Supplementary Material, Appendix \ref{sec:supp}).  These differences
are within the expected uncertainties since return value estimates from
observational time series of relatively short length (10-30 yrs) come with
large confidence intervals.  This is clearly seen for buoy 41001 east of
Cape Hatteras (\Fig{41001}), where \Uexp~estimates from ENS240 reveal yields
95\% confidence intervals of 1 m~s$^{-1}$ while ERA-I has a width of about
4 m~s$^{-1}$ and the observations (1980-2012 with some gaps) in excess of
6 m~s$^{-1}$.

%Comparison_with_earlier_studies
How do our return values compare with previous investigations?  \Udre~and
\Hdre~exhibit the same geographical features as found by \citet{caires05b}
who calibrated the earlier ERA-40 reanalysis \citep{upp05} to \textit{in situ}
measurements of \U~and \Hs.  However, their \Hr~fields are up to 6 m higher
in the storm tracks in the North Atlantic and the North Pacific compared
to what we find.  No confidence intervals are provided, but a comparison
of decadalwise return values shows approximately 10\% spread in \U~and
\Hs~in the extratropical storm tracks.  Estimates of \Ur~and \Hr~based on
aggregated altimeter measurements from satellite missions over the past 30 yrs
\citep{vinoth11,young12} also yield much higher return values for wind speed in
the extratropics. Wind speeds in excess of 52 m~s$^{-1}$ in the North Atlantic
and the North Pacific are reported \citep{vinoth11,young12} using IDM. This
is almost 50\% higher than what we find and is likely to be an artefact of
IDM, something which is also hinted at by the fact that the GP estimates
\citep{vinoth11} fail to reproduce these extremes. For \Hr~the results are
much closer to our estimates.  Since in the extratropics ENS240 compares well
with \textit{in situ} observations (both \U~and \Hs) and altimeter wind speed,
we conclude that previous estimates \citep{caires05b,vinoth11,young12}
are probably too high outside the subtropics and much too low in the
tropics and subtropics (as are our estimates).  Note that \citet{caires05b}
calibrated against 3-h averaged observations whereas \citet{vinoth11} and
\citet{young12} assumed that the altimeter observations represented 3-h
averages. This will lead to somewhat higher return values since we assume
that the values are representative of 6-h interval. However, this can only
explain a small fraction of the discrepancy (see \Fig{4h}).

We note also that the investigation of 20-yr return values of \Hs~ from
altimeter measurements by \citet{izaguirre11} yielded a positive shape
parameter (unbounded extreme values) in areas with tropical cyclones, but
a negative shape parameter (bounded extremes) in the extratropics. Their
Fig~1 yields a map which is qualitatively very similar to \Fig{drediff}a.

%Discussion_of_nonstationarity_and_return_values
\section{Discussion}
\label{sec:discussion}
Implicit in the term ``return value'' lies the tacit assumption that
the probability of non-exceedance is drawn from a long, stationary time
series \citep{col01}. In practice, observed and modeled time series vary
in length from 30-100 yrs and are thus much shorter than the period over
which stationarity is assumed, and with the exceptions of \citet{compo11},
\citet{wang12} and \citet{hersbach13}, the time series are also substantially
shorter than the 100-yr return period. Decadal variations in storminess may
affect these estimates \citep{caires05b,young12} and climate projections
of future wind and wave climate \citep{wang06,debernard08,hemer13} suggest
that nonstationary methods must be applied to assess future return values
(see \citet{kharin05}, \citet{izaguirre10} and \citet{izaguirre11} for
examples of nonstationary extreme value analysis).  There is disagreement
about the presence of trends in wave extremes in the extratropics
\citep{wang02,caires05b,young12}, but there is some evidence \citep{young12}
that over the past three decades there has been a positive trend in extreme
wind speed globally as well as in the intensity of tropical cyclones
\citep{elsner08}.  The trends in extreme wind speed \citep{young12}
are, however, based on 10-yr chunks of altimeter measurements and the
uncertainties are very high. This is one of the strengths of using large
aggregates of ensemble forecasts.  These will by construct not exhibit
long-term trends and low frequency oscillations since the initial conditions
cover a period of 9 yrs. Stationarity is thus not an issue, suggesting that
a similar approach could be used on short time slices (about 10 yrs) from
ensembles covering several decades to avoid the complicating factors of using
nonstationary methods. Candidate datasets that approach the size required
would be reforecast ensembles \citep{hagedorn12} and the recently completed
20th century model 10-member ensemble integrations \citep{hersbach13}.
For climate projections multi-model ensembles \citep{hemer13} could provide
the amount of data necessary for extreme trend estimates.

%Outro
Very large ensembles allow us to sidestep the choice of parametric distribution
as we can compute direct 100-yr non-parametric return estimates.  In principle,
by looking at forecast ranges (briefly investigated by B13) instead of
a fixed lead time it would be possible to perform a peaks-over-threshold
analysis and increase the dataset by an order of magnitude. This would also
remove any ambiguity concerning the representative interval of forecasts
as the forecast range would determine the equivalent length of the dataset.
Candidates for such an analysis would be for example the ECMWF monthly forecast
system \citep{vitart08}, the extended forecast range of IFS from day 10 to
day 15 and the seasonal forecast system \citep{stockdale11,molteni11}.

Using very large ensembles at advanced lead times is a novel method previously
only explored for \Hs~in the Northeast Atlantic (see B13).  The method,
like any method employing modeled fields, is sensitive to tail biases in the
numerical model \citep{magnusson14}.  These biases must be assessed for each
new parameter, but in our case we find that they are moderate outside areas
with tropical cyclones and significantly lower than those found for ERA-I. On
the other hand, the unsystematic error associated with the parameter fitting
to an extreme value distribution \citep{col01} is directly related to the
size of the dataset and thus benefits immediately from an increase in size.
The method is general and holds the promise of making much larger
datasets available for a wide range of oceanographic, hydrological and
meteorological parameters that are routinely forecast and archived.

\appendix
%%% Methods summary
\section{Methods Summary}
\label{sec:methods}
The ENS240 forecasts were interpolated onto a regular $1^\circ \times 1^\circ$
grid. All +240-h forecasts (two per day, 00 and 12 UTC, 51 ensemble members
in each) between 26 March 2003 and 25 March 2012 were used.  ERA-I analyses
(1979-2012) were interpolated onto the same grid.  The current spectral
truncation of ENS240 is T639 for the atmospheric model, corresponding
to approximately 32 km, whereas the wave model ECWAM is run at approximately 55
km. Previous model cycles had coarser resolution, see Supplementary Material,
Appendix \ref{sec:supp} and B13.  The 10-m neutral wind speed was extracted
for both ENS240 and ERA-I. This is the field used to force the wave model
and is thus consistent with the \Hs~fields investigated. Grid points that
are ice-covered in more than 20\% of the forecasts were censored from the
analysis of \Hs. No such censoring is required for \U.

We compared GP and EXP estimates and decided to use EXP for our comparison
of ENS240 return values with ERA-I and \textit{in situ} observations
(see Supplementary Material, Appendix \ref{sec:supp}). The threshold was set to the 1000th highest
forecast, corresponding to $P_{99.7}$. This choice (see Supplementary
Material, Appendix \ref{sec:supp}) was made after investigating the stability of EXP and GP estimates
as a function of the threshold. Since ensemble forecasts are assumed
uncorrelated, all values exceeding the threshold were used. For ERA-I and
the \textit{in situ} observations, the threshold was also set to $P_{99.7}$
and a POT technique applied where peaks must be separated by 48~h to ensure
all entries are independent \citep{lop00,col01}. The distribution fitting
was done with the maximum likelihood method and confidence intervals were
estimated by bootstrapping. We found that 500 resamples were sufficient
to yield a stable mean and a stable 95\% confidence interval (see Supplementary
Material, Appendix \ref{sec:supp}). The threshold was set high enough to avoid summertime values. This
means that in locations with strong seasonality in wind speed and direction
our estimates should be interpreted as wintertime return values. We make no
attempt to stratify the dataset by season or direction.

Under the assumption that the collection of +240-h forecast members are
equivalent to a temporal period (each forecast representative of a 6-h
interval, see Supplementary Material, Appendix \ref{sec:supp}) we may convert our collection of
ensemble forecasts into an equivalent time series.  The two daily ensembles
of +240-h forecasts from 50 perturbed ensemble members plus the unperturbed
control member aggregated over 9 yrs can be considered to represent
\begin{equation}
    9 \,\mathrm{yr} \times 365.25 \,\frac{\mathrm{days}}{\mathrm{yr}} \times
    2 \,\frac{\mathrm{forecasts}}{\mathrm{day}} \times 6 \, \mathrm{h} \times 
    51\, \mathrm{mem} 
  = 2,011,797 \,\mathrm{h} = 229.5 \,\mathrm{yr}.
  \label{eq:size}
\end{equation}
We can therefore make non-parametric direct return estimates from the ensemble
of the 100-yr return value, $x_{100}^\mathrm{DRE}$, without invoking a
theoretical extreme value distribution (see Supplementary Material, Appendix
\ref{sec:supp}). 

The assumption that the modeled wind and wave fields are representative of
a 6-h interval is based on considerations of the horizontal model resolution
\citep{bidlot02}.  To test the sensitivity of the return value estimates to
this assumption we have reduced the interval to 4~h, which is close to the
averaging period used for the observations ($\pm 2$~h). This reduction leads
to an increase of at most 0.6~m for \Hr~and 1.2 m~s$^{-1}$ for \Ur~ using
the EXP distribution (\Fig{4h}). This suggests that the method
is reasonably insensitive to the choice of interval, but a little conservative.

All return estimates were computed marginally, i.e., per grid
point.  After that all fields were smoothed using a $5^\circ\times 5^\circ$
two-dimensional box-car filter after applying a smaller ($3^\circ\times
3^\circ$) MAX filter which selects the highest value from the nearest
neighbours. This counters the tendency of the smoothing to lower the highest
return values.

\section{Supplementary Material}
\label{sec:supp}
The Supplementary Material is only available online in the article in
\textit{Geophys Res Lett}. Here, we include it as Appendix \ref{sec:supp}. The
figures thus form part of the manuscript and are numbered differently from the
final version. 

\subsection{The ECMWF Integrated Forecast System (IFS) Ensemble}
Since March 2003 IFS has been run twice daily to 10 days with 51 ensemble
members \citep{buizza07}. As discussed by \citet{bre13b}
(see their Fig 1), the resolution and the model physics of both the atmosphere
and the wave components have undergone many revisions. The main changes in
terms of resolution were upgrades to T399 from T255 in February 2006 and a
further upgrade to T639 in February 2010. WAM followed this general increase in
resolution with increases in spatial resolution from $1.0^\circ$ to $0.5^\circ$
in February 2010. The same subgrid scheme for unresolved bathymetry as was
used for the ERA-I implementation was implemented in March 2004.  For this
investigation we have interpolated to the same regular longitude-latitude
grid of $1.0^\circ$ as has been employed for the investigation of the ERA-I
reanalysis.

\subsection{ERA-Interim}
ERA-Interim (ERA-I) is a global coupled atmosphere-wave ocean reanalysis
starting in 1979 \citep{simmons07,uppala08,dee11} based on Cycle 31r2 of IFS. The archive is extended in
time regulary to be up to date. In this study the period covered is 1979-2012
(34 yrs). The archive has a resolution of $1.0^\circ$ for the wave model at
the equator. An irregular latitude-longitude grid ensures relative constancy
in grid resolution towards the poles.  The atmospheric spectral truncation is
T255 ($\simeq 79$ km), but atmospheric parameters are archived on a Gaussian
grid of approximately $0.75^\circ$. Both significant wave height and wind
speed are interpolated to a regular longitude-latitude grid of $1.0^\circ$
resolution for this study.  ERA-I employs a four-dimensional variational
assimilation scheme. The ERA-I WAM model is coupled to the atmospheric model
through exchange of the Charnock parameter \citep{janssen89,janssen91,jan04}. The
model incorporates shallow-water effects important in areas like the southern
North Sea \citep{kom94}.  ERA-I uses a subgrid scheme to represent the downstream
impact of unresolved islands \citep{bidlot12}.

\subsection{Model Tail Bias Assessed with In Situ and Altimeter Observations }
To assess the tail behavior of the data sets we compare against 24
quality-controlled time series \citep{bidlot02} of \emph{in situ} measurements
of 10-m wind speed, \U, and significant wave height, \Hs, in the northern
hemisphere \citep{bidlot02}.  To make the observations comparable with model
output the observations are averaged over 5~h centered on the synoptic times.
ENS240 wave fields are relatively coarse (approximately 50 km for T639
resolution) and should be considered representative of intervals of 4-6~h.
We find good agreement at the 99.7\% percentile ($P_{99.7}$) level for both
\U~and \Hs\,[see \Fig{P99}, locations indicated in \Fig{envisat}].  ENVISAT RA2
altimeter observations of surface wind speed covering the period 2002-2012
were averaged into along-track ``super-observations'' \citep{janssen07}
of a resolution similar to the WAM model grid \citep{abdalla04, dragani14}.
This procedure makes data and model values more comparable and serves the same
purpose as the temporal averaging of \emph{in situ} observations.  \Fig{envisat}
shows that ERA-I is typically biased low by more than $-1.5$ m s$^{-1}$ in
certain regions and about $-0.5$ m s$^{-1}$ globally.  This is in contrast
to ENS240 which is virtually unbiased when averaged globally and rarely
deviates more than $\pm 0.5$ m~s$^{-1}$ locally.

As a further test of the tail behavior we have compared return values computed
with the exponential distribution from \emph{in situ} measurements (identical
to how the ERA-I estimates were computed) to the ENS return estimates. As
\Fig{reg100} shows, the estimates match well, with less than 15\% deviation
in both wind speed and significant wave height except for \emph{in situ} locations
in the Gulf of Mexico exposed to tropical cyclones. As the time series for
the \emph{in situ} observations range from 10-30 yrs, there is large uncertainty in
the estimates based on the observations.  We would like to emphasize, as we
do in the article, that tail (upper-percentile) biases may seriously affect
the return value estimates. However, this is a problem always present when
using model data for extreme value estimation, and is if anything less of a
problem with the ENS240 dataset than with ERA-I, which tends to be biased low.

\subsection{Estimating Return Values From Ensembles}
Breivik \emph{et al} \citep{bre13b} laid down four criteria for using ensembles
for extreme value estimation, restated slightly here as
\begin{enumerate}
  \item No significant correlation between ensemble members at advanced lead 
  times 
  \item The model climatology is comparable to the observed climatology
  distribution
  \item No spurious trend due to model updates
  \item Forecasts are representative of a time interval (e.g. 6 h) 
\end{enumerate}
Comparison with reforecast ensembles \citep{hagedorn12} found (see
\citet{bre13b}) no spurious trend in the mean and the variance of \Hs~
due to model upgrades.

As outlined by \cite{bre13b}, converting $M$ ensemble forecasts with $N$
ensemble members each into the equivalent of a time period is necessary to
go from probability of exceedance to return periods. We have assumed that
each forecast represents a six-hour interval. The justification for this is
firstly that $\Delta t = 6\,\mathrm{h}$ matches the temporal resolution of
ERA-I, hence making it easy to compare the results. Secondly, model fields are
smoothly varying in time, making them representative of averages over typically
4-6 h at the resolution of ERA-I and ENS240. This allows us to treat the
collection of ensemble forecasts as an equivalent time period $T_\mathrm{eq}
= M N \Delta t$. This averaging period coincides well with that used for
the smoothing of the wave and wind observations \citep{bidlot02}, which is
$\pm 2$ h. However, there is some arbitrariness in our choice of $\Delta t =
6\,\mathrm{h}$. 

Parametric return estimates from ENS240 were found using the GP distribution
for values exceeding a threshold $u$ such that $y = X_i - u$, $y > 0$. The
distribution can be written \citep{col01} as
\begin{equation}
   H(y) = 1 - \left(1+\frac{\xi y}{\tilde{\sigma}}^{-1/\xi}\right).
   \label{eq:gp}
\end{equation}
Here $\xi$ is the shape parameter and $\tilde{\sigma}$ is a scale parameter.
In the limit as $\xi \to 0$, $H(y) = 1 - \exp(-y/\tilde{\sigma})$, i.e.,
an exponential distribution (EXP) with parameter $1/\tilde{\sigma}$.
We follow \citet{col01} and estimate return values $r_T$ with
  \begin{equation}
    r_T=u+\frac{\tilde{\sigma}}{\xi}\left[ \left( T n_\mathrm{y} \zeta_{u} 
                                     \right)^{\xi}-1\right],\quad \xi \neq 0
  \end{equation}
  \begin{equation}
    r_{T}=u+\tilde{\sigma} \text{log}\left(T n_\mathrm{y}\zeta_{u}\right),\quad \xi=0
  \end{equation}
  \begin{equation}
    \hat{\zeta}_{n}=\frac{k}{n}.
  \end{equation}
Here $n$ represents the total number of data, $n_\mathrm{y}$ is the total
number of data per year, $k$ is total number of peaks  above the threshold
$u$, $\hat{\zeta}_{n}$ is an estimate of the proportion of peaks above the
threshold $u$, $T$ is the $T$-year return period, and $\tilde{\sigma}$ and $\xi$ are
the scale and shape parameter estimates obtained with the Maximum Likelihood
Method \citep{col01}.  Assuming the reference period to be 6 hours
  \begin{equation}
    n_\mathrm{y} = 365.25\,\text{days}\times 4\,\text{entries per
    day}=1461\,\text{entries}.
  \end{equation} 
To investigate the impact of our choice of reference interval we 
recomput the return values using $\Delta t = 4\,\mathrm{h}$,
  \begin{equation}
    n_\mathrm{y} = 365.25\,\text{days}\times 6\,\text{entries per
    day}=2191.5\,\text{entries}.
  \end{equation}
In the extratropics the impact is modest (\Fig{4h}) with at most an increase
in \Ur~ of 0.6 m~s$^{-1}$ and 0.5 m for \Hr~ for EXP estimates. Areas
prone to tropical cyclones east of Madagascar and in the South China Sea
see an increase in \Ur~ of up to 1.2 m~s$^{-1}$.  For the GP distribution
(not shown) the discrepancy in $H_{100}$ will be larger in areas of $\xi>0$,
i.e. in areas visited by tropical cyclones, and smaller in areas with $\xi<0$,
i.e. in the extratropics, see \citet{izaguirre11}.

It is also of interest to look at the absolute maxima obtained from the
ensemble. \Fig{max} shows that \U~ comes close to 50 m~s$^{-1}$ and \Hs~
reaches almost 25 m. The filtering effect of ocean waves is evident from
the fact that the max field is spatially much smoother for \Hs~ than for \U.

\subsection{Direct Return Estimates from Order Statistics}
Consider a stationary time series of length $NT$ where $T$ is the return
period.  By definition the $N$ highest entries $x_{(i)}$ are greater than
the return value $r_T$. Thus each of the $N$ periods contains on average one
entry at least as large as $r_T$. These $N$ highest entries are independent
and spread randomly over the entire domain $NT$.  Let $\tau T$ represent the
period of interest. Now $\tau$ is time measured in units of the return period.
There is a probability $1-\tau/N$ that a randomly chosen interval of length
$\tau T$ does not contain a given entry $x_{(i)}$.  As $N$ becomes infinitely
large the probability that the interval contains \emph{neither} of the $N$
highest values is [\cite{abr72}, Eq~(4.2.21)]
\begin{equation}
   P(x_\tau < r_T) = \lim_{N\to\infty} (1-\frac{\tau}{N})^N = \mathrm{e}^{-\tau}.
   \label{eq:neither}
\end{equation}
Here $x_\tau$ denotes the highest entry in $\tau T$.  This result is
independent of the underlying probability distribution as only the rank of
the highest ${x_{(i)}}$ is of interest.  It is now clear that some care has to
be taken when interpreting the upper percentiles of even quite large datasets
since according to \Eq{neither} the probability of exceeding the return value
$r_T$ in any given return period $\tau = 1$ is $1-1/\mathrm{e} \approx 0.63$.
Conversely, there is still a certain probability (${\approx}0.1$) that the
100-yr return value does not appear in our dataset of magnitude equivalent
to 229 yrs ($\tau = 2.29$).

Our direct 100-yr estimate (DRE) taken from the order statistics of a
dataset of this length is a linear interpolation between the second and third
highest values [denoted $x_{(2)}$ and $x_{(3)}$] in the ordered sequence. The
interpolation weight $w$ is found through the relation
\begin{equation}
     w\frac{2}{229} + (1-w)\frac{3}{229} = \frac{1}{100},
   \label{eq:weights}
\end{equation} 
where the fractions represent frequencies of occurrence of the second and third
highest events as well as the 100-yr return event. Thus
\begin{equation}
   r_{100}^\mathrm{DRE} = 0.67x_{(2)}+0.33x_{(3)}.
   \label{eq:dre}
\end{equation}

\subsection{Bootstrapping Confidence Intervals}
Confidence intervals have been computed for ENS240 Generalized Pareto (GP)
and exponential (EXP) distributions. It is interesting to compare the width
of confidence intervals with those derived from ERA-I.  With ERA-I we must
perform a peaks-over-threshold (POT) selection of storm events, which is
not applicable to the uncorrelated ENS240 data set.  We have chosen in the
main article to compare EXP return value estimates and used bootstraps of
500 resamples for the computation of the confidence intervals.  The same
relative differences are found for GP confidence intervals, although much
wider in absolute terms (\Fig{erai_gp}b).

The bootstrap size must be large enough to provide a stable estimate of
the return value but also of the confidence intervals. \citet{efron87}
demonstrated that in general bias-corrected accelerated (BCa) confidence
intervals are preferable. We find that beyond a size of 500 the bootstrap
yields very stable estimates of confidence intervals and average return values.
\Fig{bootsize} demonstrates this clearly for the case with GP estimates (Panel
a) and EXP estimates (Panel b) for a location in the Norwegian Sea.

\subsection{Threshold Selection}
The threshold should ideally be chosen individually for each location
\citep{aar12}, but this is clearly not feasible with a global dataset. We
have chosen the level 99.7\% after studying the behavior of both the GP
and EXP estimates for a selection of locations. \Fig{threshold} shows the
behavior of the GP (blue) and EXP (red) estimates. The two estimates start
to converge above 99\% and are quite stable around 99.7\%. This behavior
with some variation is found throughout the extratropics.  Selecting a
threshold of 99.7\% also means that the dataset contains only winter time
storms in the extratropics and as such the estimates should clearly be seen
as winter time return estimates. We make (as we stress in the main article)
no attempt at stratifying our data by season or wind direction.
To further investigate the impact of our choice of threshold we plot the
global difference map between EXP estimates at $P_{99.1}$ and $P_{99.7}$ in
\Fig{threshmap}. As can be seen the estimates are not extremely sensitive to
the choice of threshold except in areas with tropical cyclones. Given what we
know from \Fig{threshold} about the behavior of the return values for locations
in the extratropics the threshold $P_{99.7}$ seems a reasonable choice.

\subsection{Influence of the Shape Parameter}
The shape parameter $\xi$ [see \Eq{size}] is highly sensitive to
the tail of the empirical distribution function. This can lead to spatially
highly variable fields of return values. With data sets shorter than the
return period, which is normally the case with observational records and
model reanalyses, it may be difficult to find the shape parameter which best
represents the conditions in a large geographical domain. Comparing the EXP
distribution ($\xi = 0$, \Fig{dre_v_exp}) and the GP distribution ($\xi
\neq 0$, see \Fig{drediff}) with the direct return estimates (DRE) reveals
that for wind speed outside the tropics both EXP and GP agree well with
DRE. However, in areas with tropical cyclones EXP tends to underestimate while
GP yields return values substantially above DRE.  GP may be a sensible choice
for winds since our model results will tend to be biased low in the tropics as
IFS is unable to capture the severity of tropical cyclones \citep{magnusson14}.
For \Hr~ the picture is not quite so clear (contrast \Fig{drediff}b 
with Panel b of \Fig{dre_v_exp}), but the same general tendency
for EXP to underestimate in areas with tropical cyclones lead us to conclude
that GP is a better choice.

The comparison with ERA-I return value estimates was done using the EXP
distribution (see Figs~\ref{fig:U100}b and \ref{fig:h100}b). The reason
is that the GP distribution shows erratic behavior in areas with tropical
cyclones with very large confidence intervals (see \Fig{erai_gp}).

\section*{Acknowledgment}
This work has been supported by the Research Council of Norway
 through the project ``Wave Ensemble Prediction for Offshore Operations''
 (WEPO, grant no 200641) and through the European Union FP7 project MyWave
 (grant no 284455).  SA was funded by the European Space Agency (contract
 21519/08/I-OL, ``Technical support for global validation of Envisat data
 products'').  This study has also been part of a PhD program for OJA,
 partially funded by the Norwegian Centre for Offshore Wind Energy (NORCOWE).
 All datasets presented in this study are archived in ECMWF's MARS database.

\clearpage

\bibliography{./database,./references} % for private compilation

\begin{figure}[h]
\begin{center}
(a)\includegraphics[scale=0.7]{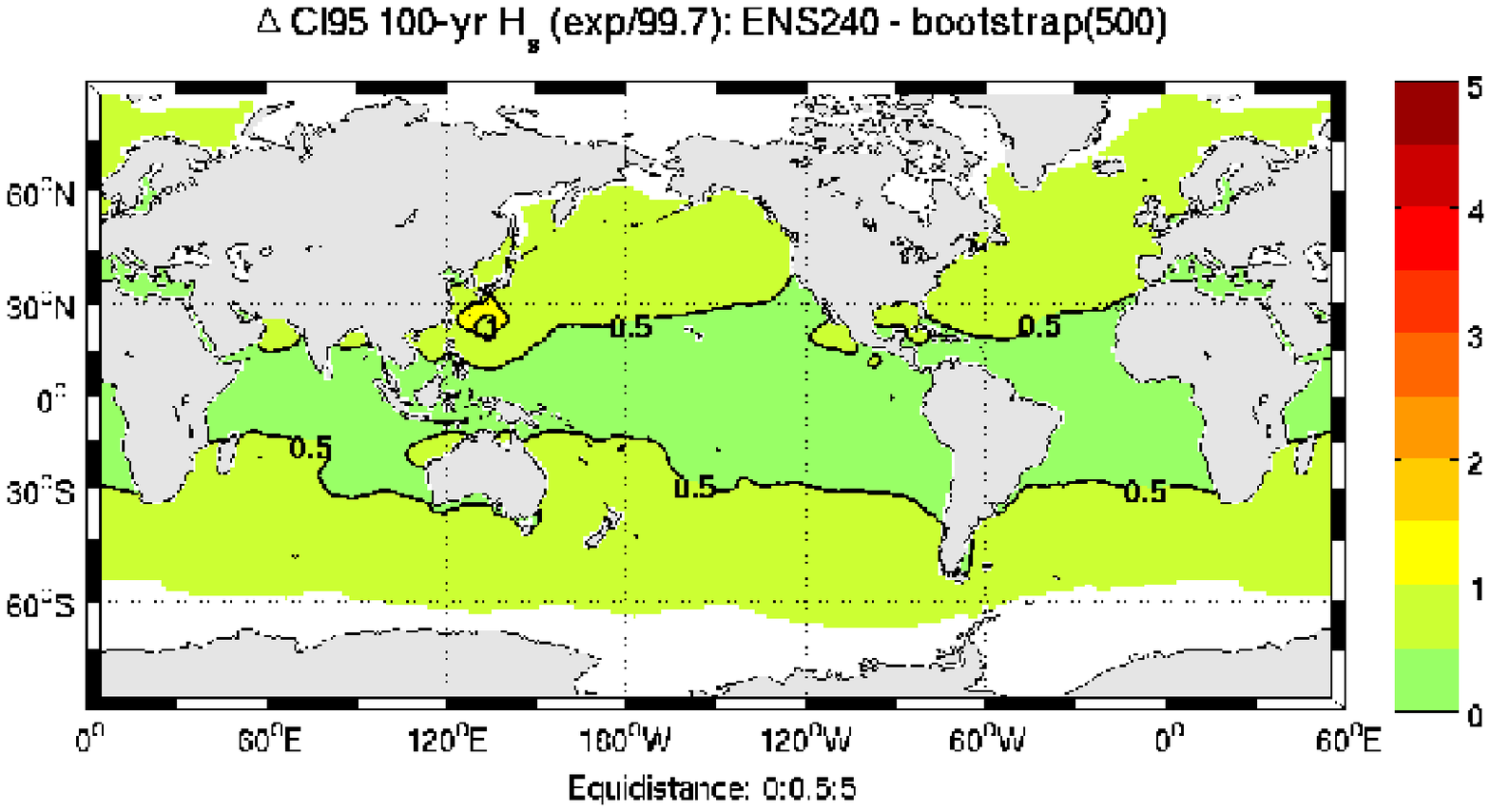}\\
(b)\includegraphics[scale=0.7]{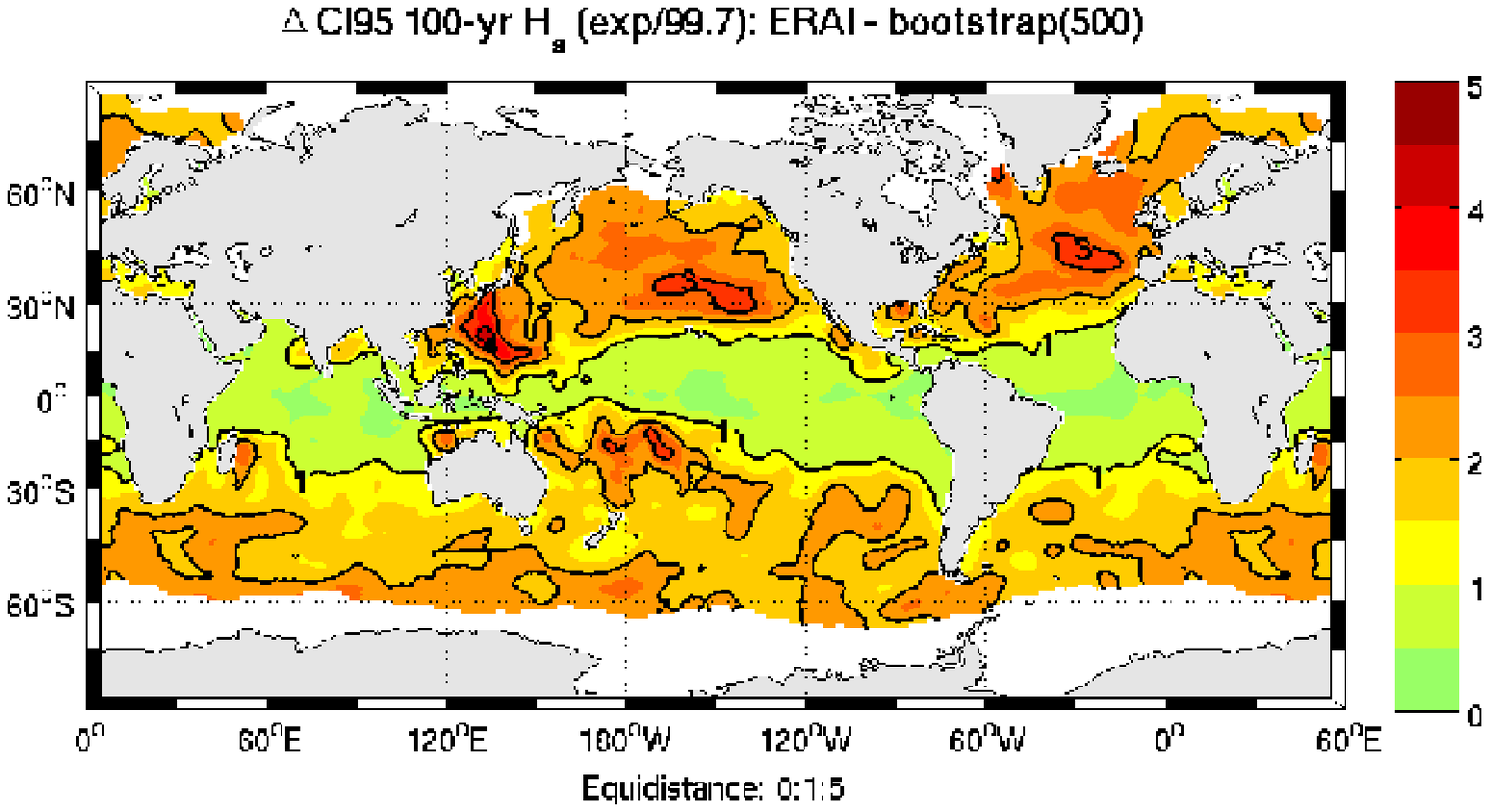}
\caption{
ENS240 v ERA-I width of \Hexp~ confidence intervals.  Panel a: Width of
the ENS240 \Hexp~ 95\% confidence interval. All data exceeding $P_{99.7}$
were used.  Panel b: Same as Panel a but for ERA-I. A peaks-over-threshold
method was used to select data exceeding a threshold of $P_{99.7}$.}
\label{fig:hexp_ci95_v_erai} 
\end{center} 
\end{figure}

%%% Supplementary figs below
\begin{figure}[h]
\begin{center}
(a)\includegraphics[scale=0.5]{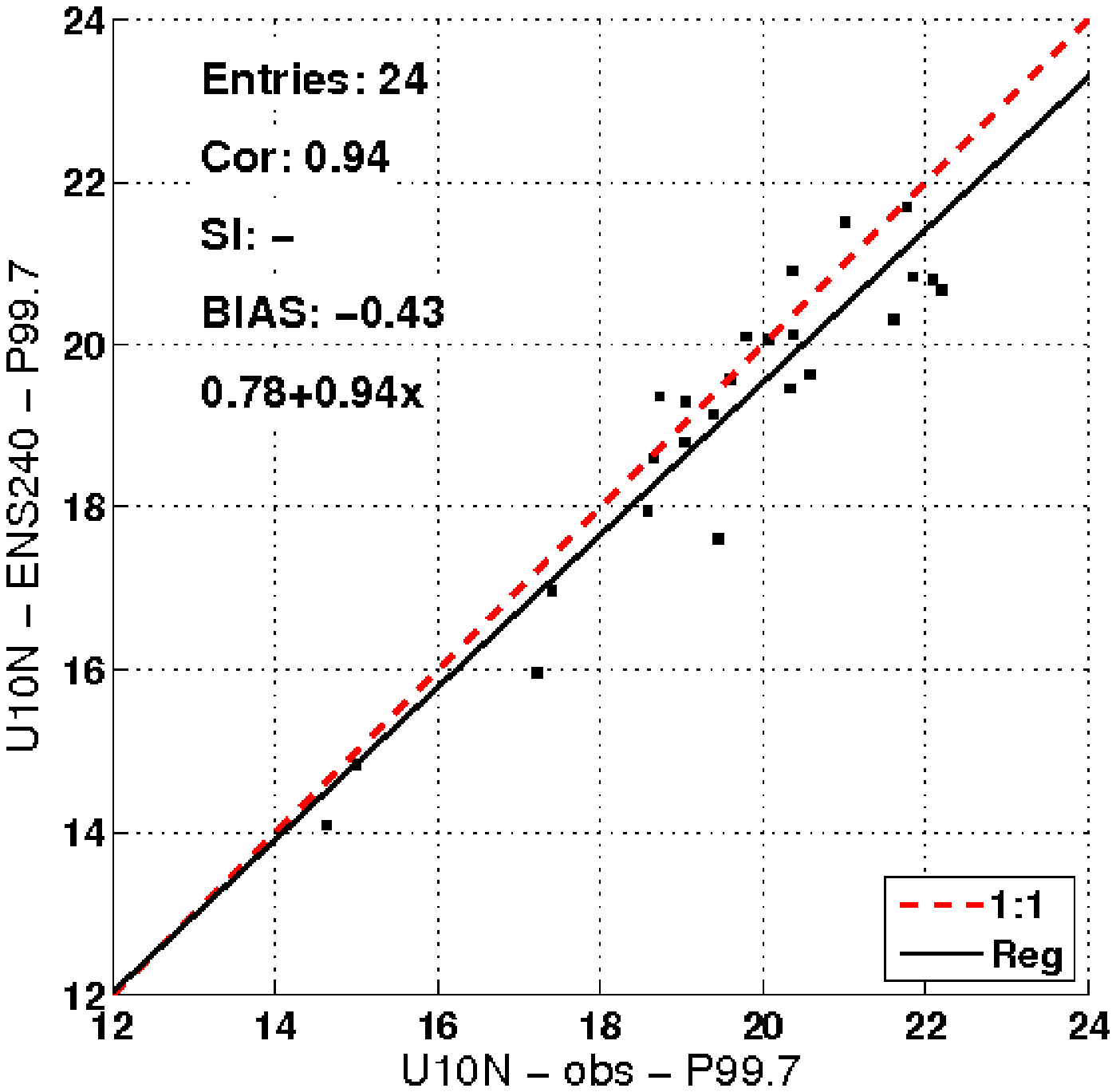}\\
(b)\includegraphics[scale=0.5]{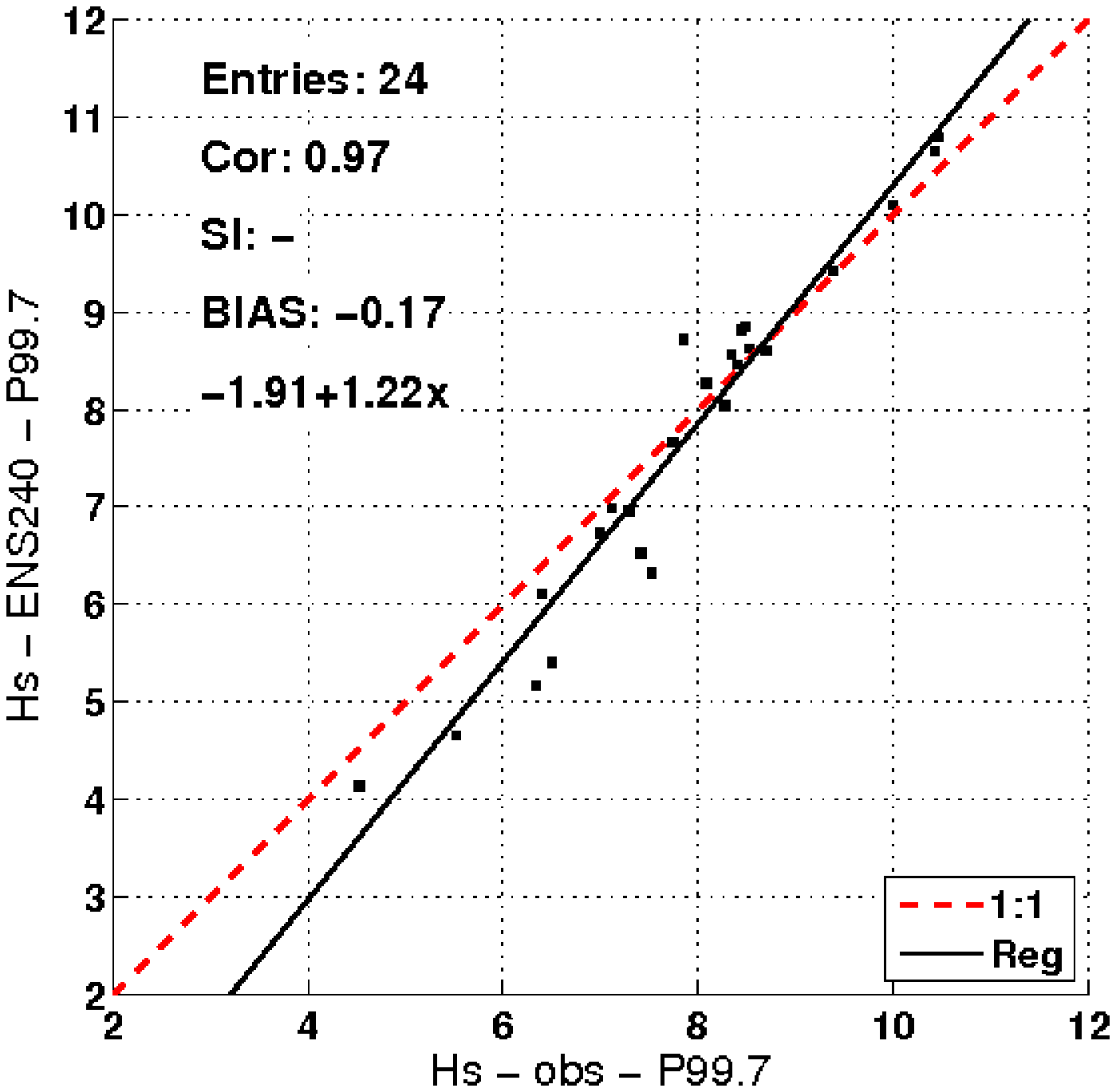}\\
\caption{
Panel a: Observed \emph{v} modeled 99.7 percentiles ($P_{99.7}$) of 10-m neutral wind 
speed [m s$^{-1}$]. Panel b: As panel a, but for significant wave height [m]. 
}
\label{fig:P99} 
\end{center} 
\end{figure}

\begin{figure}[h]
\begin{center}
(a)\includegraphics[scale=0.7]{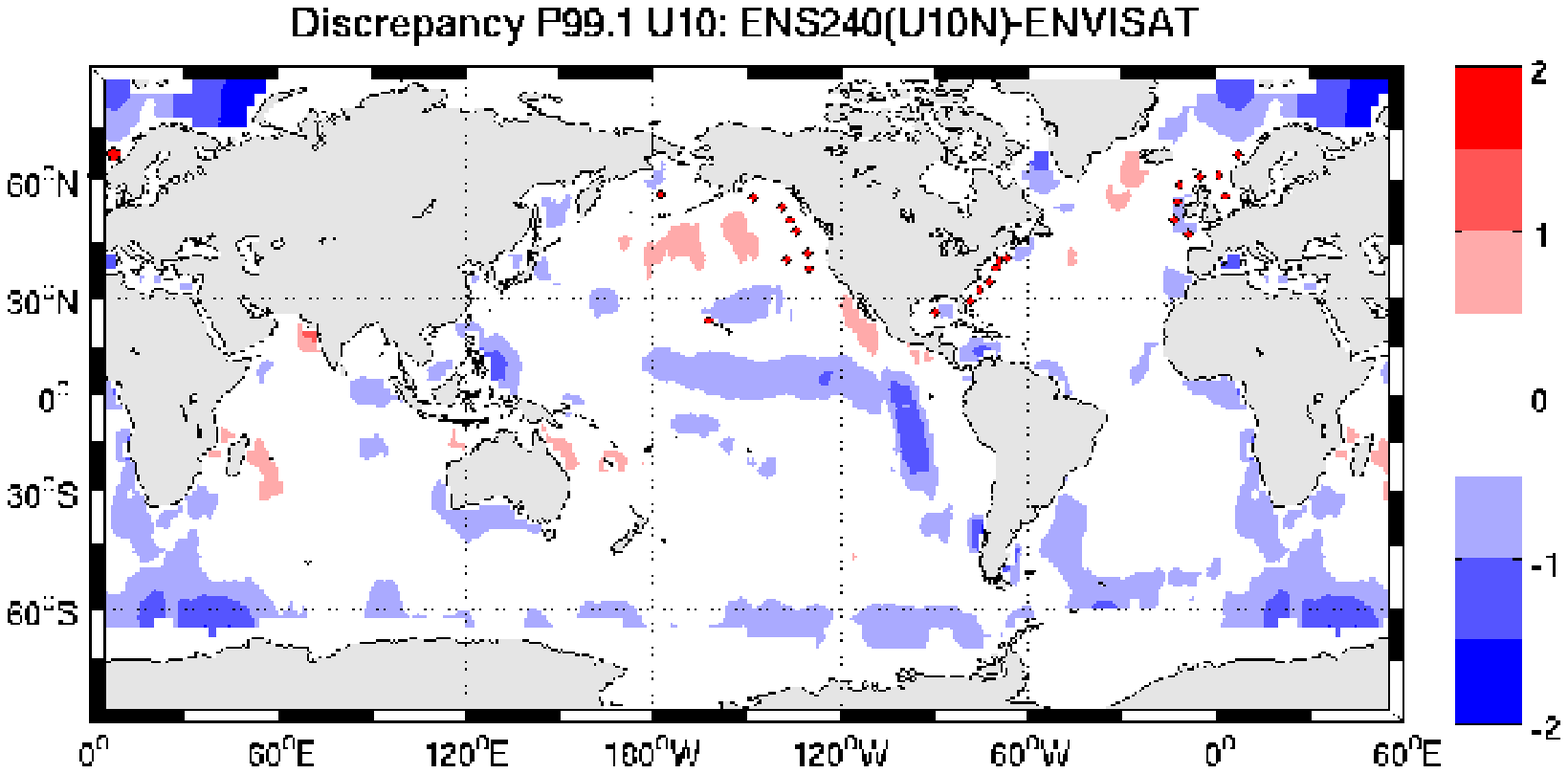}\\
(b)\includegraphics[scale=0.7]{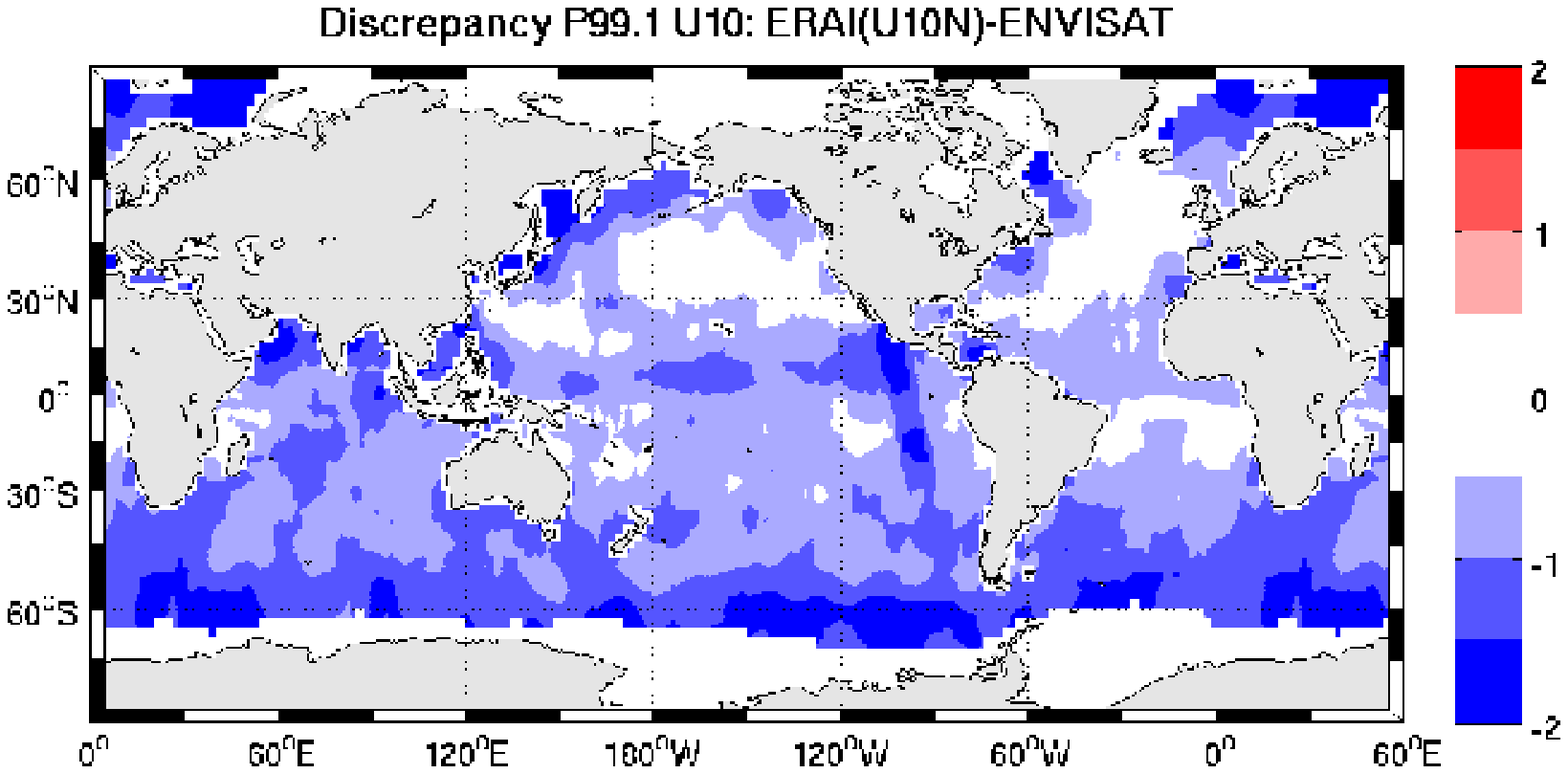}
\caption{
Panel a: The difference between the ENS240 neutral 10-m wind speed and
ENVISAT altimeter 99.1 percentile ($P_{99.1}$) (2002-2012) [m s$^{-1}$].
The 24 locations with \emph{in situ} wind and wave observations are shown
as red diamonds.  Panel b: Difference between ERA-I and ENVISAT wind speed.
The differences between ERA-I and ENVISAT at $P_{99.1}$ are generally larger
(ERA-I biased low) than for ENS240.  }
\label{fig:envisat} 
\end{center} 
\end{figure}

\begin{figure}[h]
\begin{center}
(a)\includegraphics[scale=0.5]{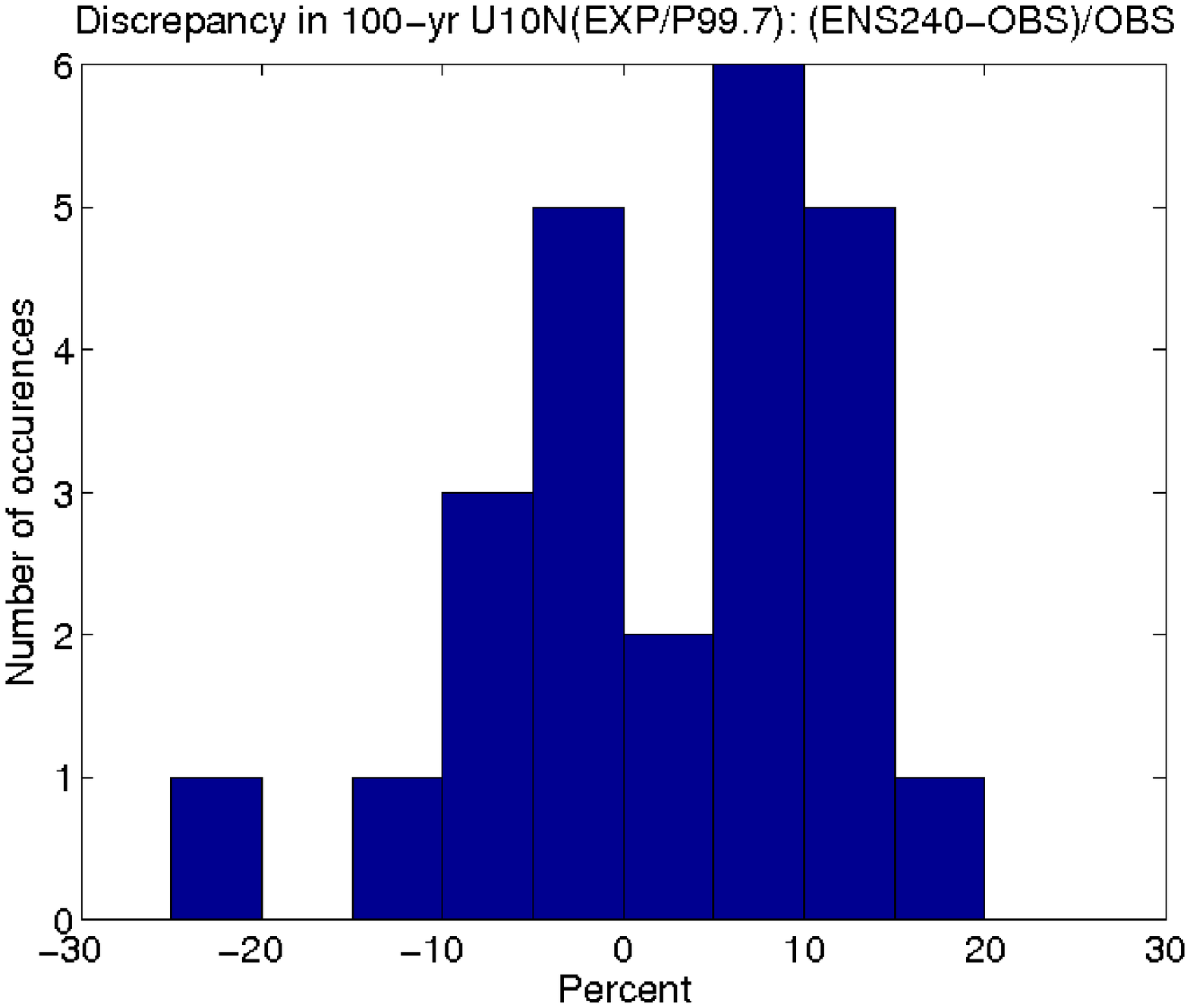}\\
(b)\includegraphics[scale=0.5]{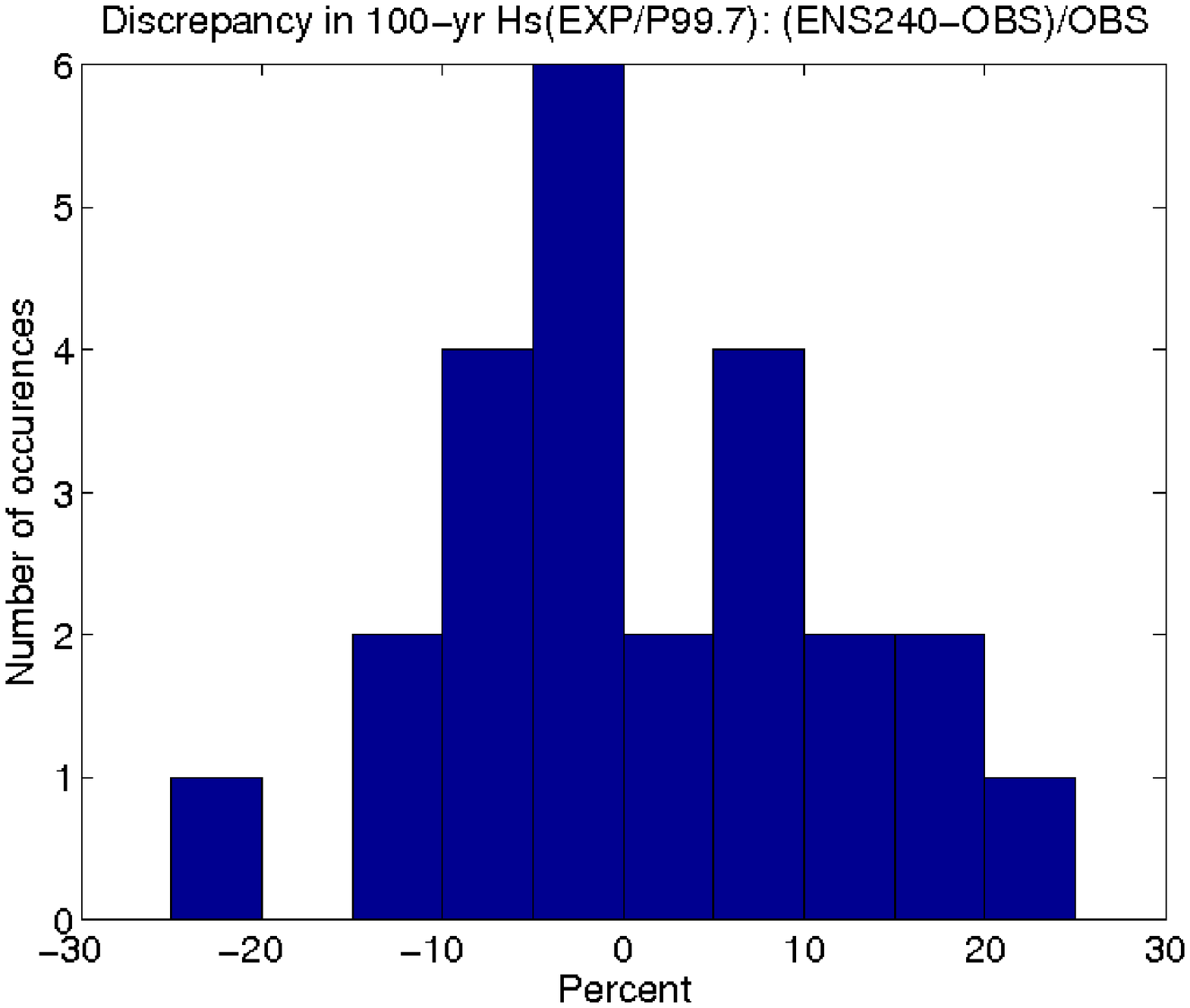}
\caption{
Panel a: Relative difference [\%] of 100-yr return values computed from ENS240
compared with return values computed from \emph{in situ} observations for
10-m neutral wind speed. Panel b: As panel a, but for significant
wave height.
}
\label{fig:reg100} 
\end{center} 
\end{figure}

\begin{figure}[h]
\begin{center}
(a)\includegraphics[scale=0.7]{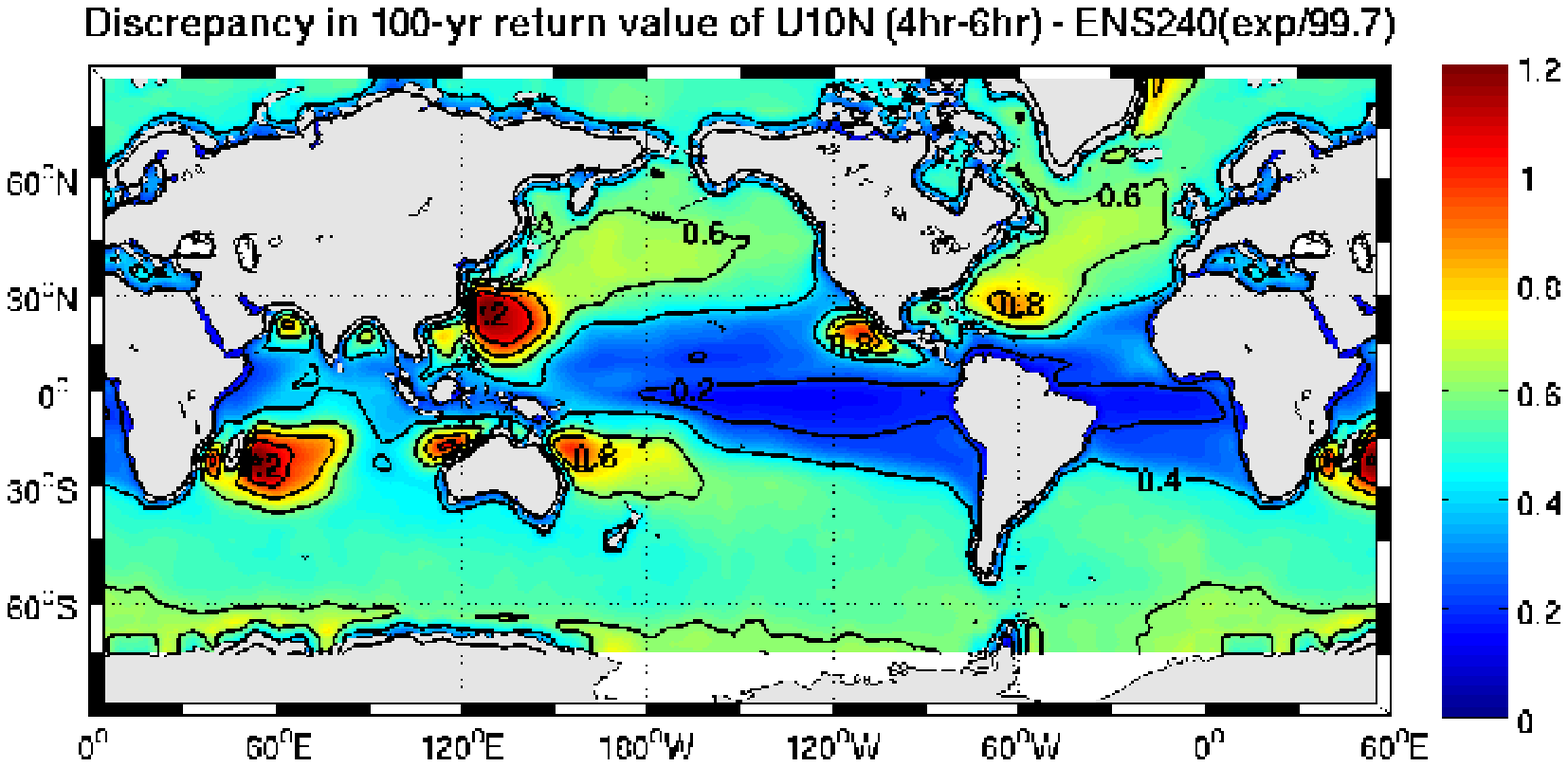}\\
(b)\includegraphics[scale=0.7]{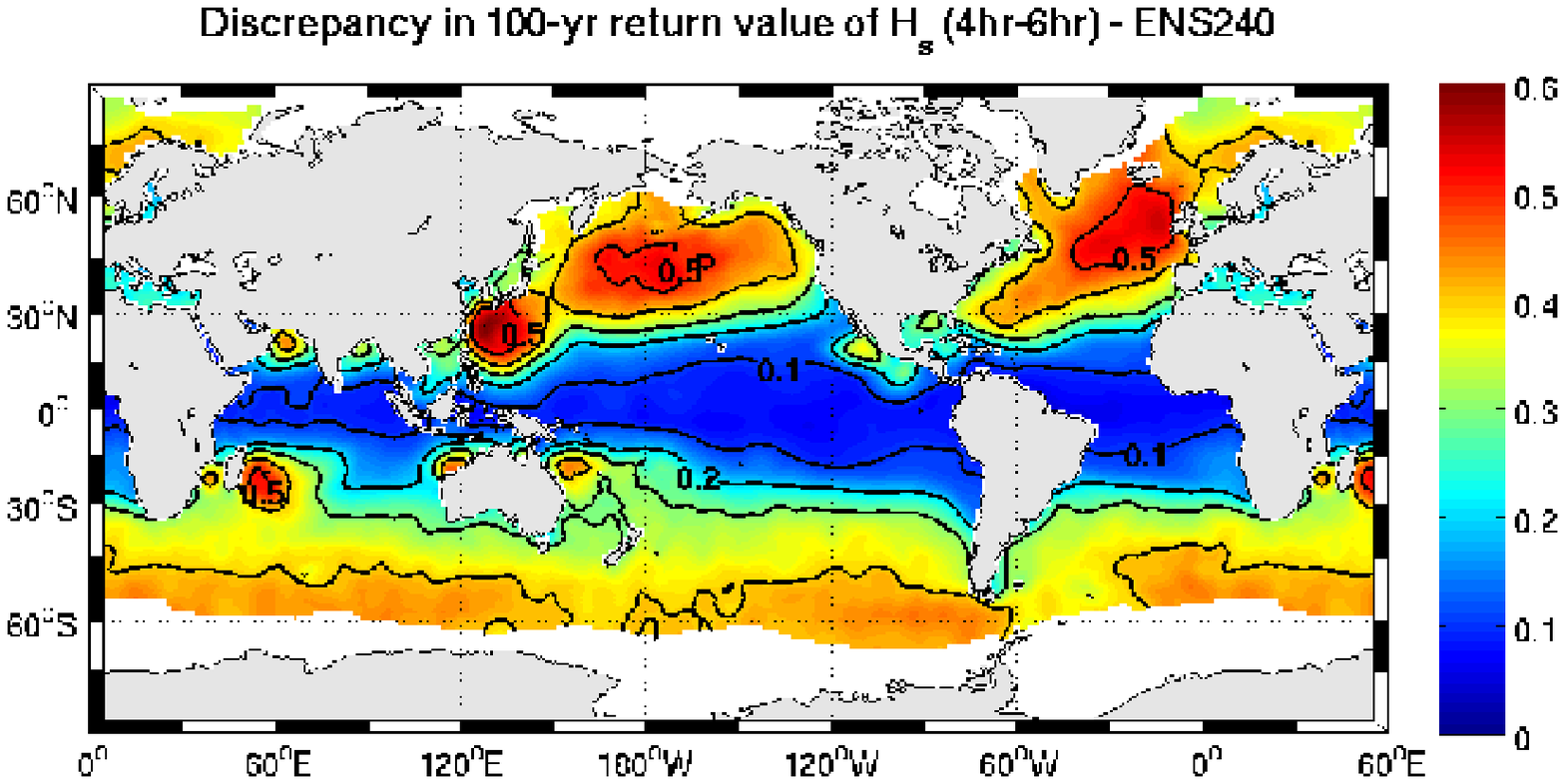}
\caption{
Panel a: The difference between ENS240 \Ur~ assuming a four-hour interval and
assuming a six-hour interval. 
Panel b: Same as panel a but for \Hr~ significant wave height.
}
\label{fig:4h} 
\end{center} 
\end{figure}

\begin{figure}[h]
\begin{center}
(a)\includegraphics[scale=0.5]{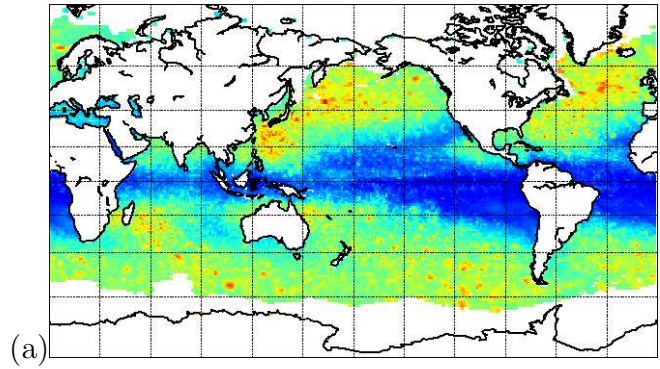}\\
\vspace{24mm}
(b)\includegraphics[scale=0.5]{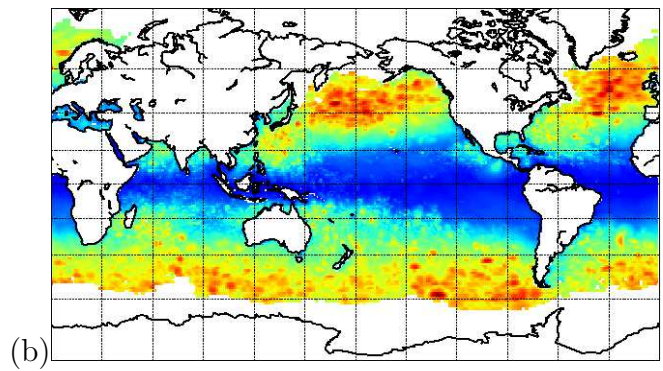}
\vspace{7mm}
\caption{
Panel a: The max field obtained from 9 yrs of archived ENS240 \U~ fields.
The highest value is 49.2 m~s$^{-1}$.
Panel b: Same as panel a for \Hs. The highest value is 24.5 m. 
}
\label{fig:max} 
\end{center} 
\end{figure}

\begin{figure}[h]
\begin{center}
(a)\includegraphics[scale=0.7]{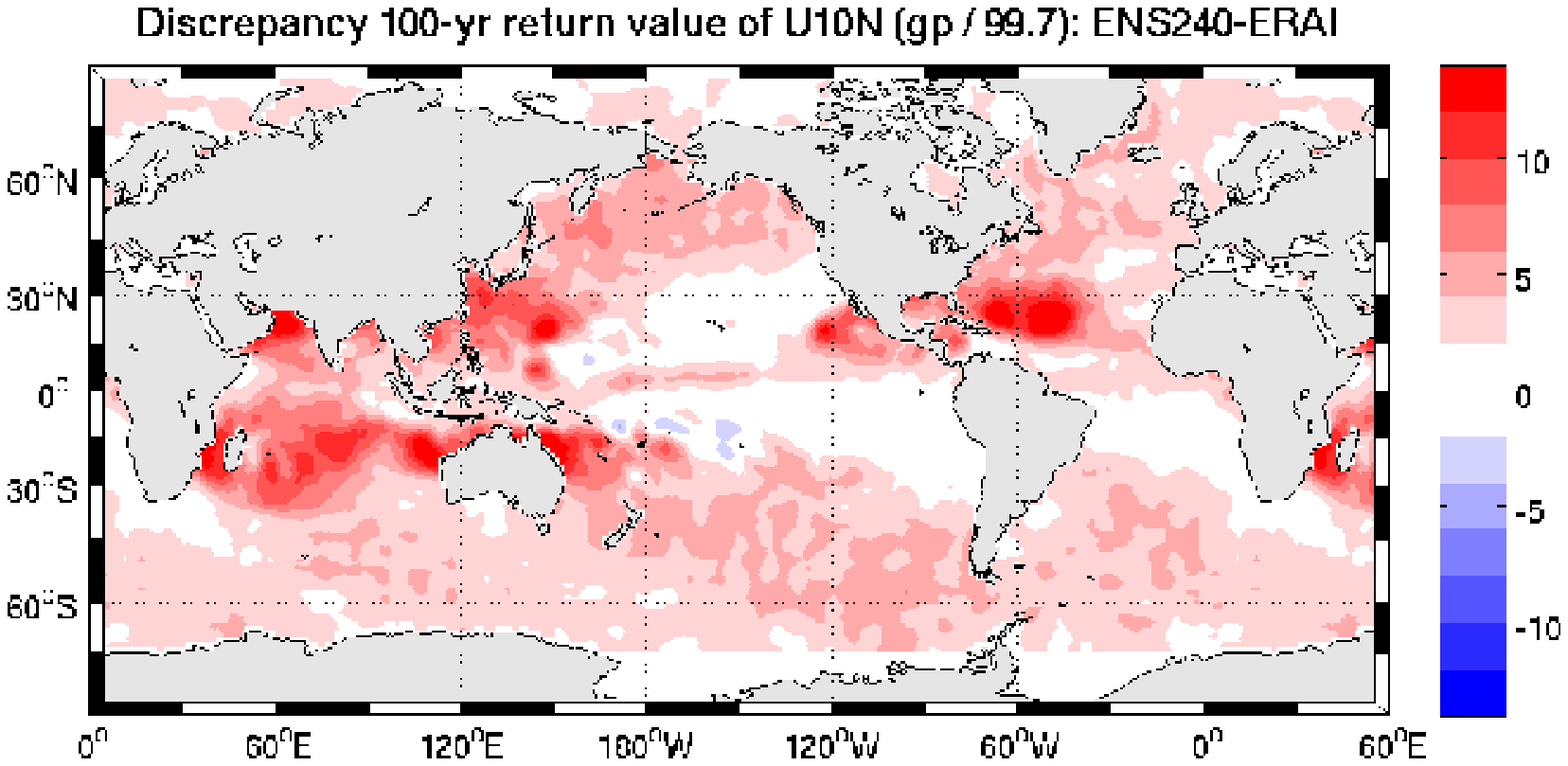}\\
(b)\includegraphics[scale=0.7]{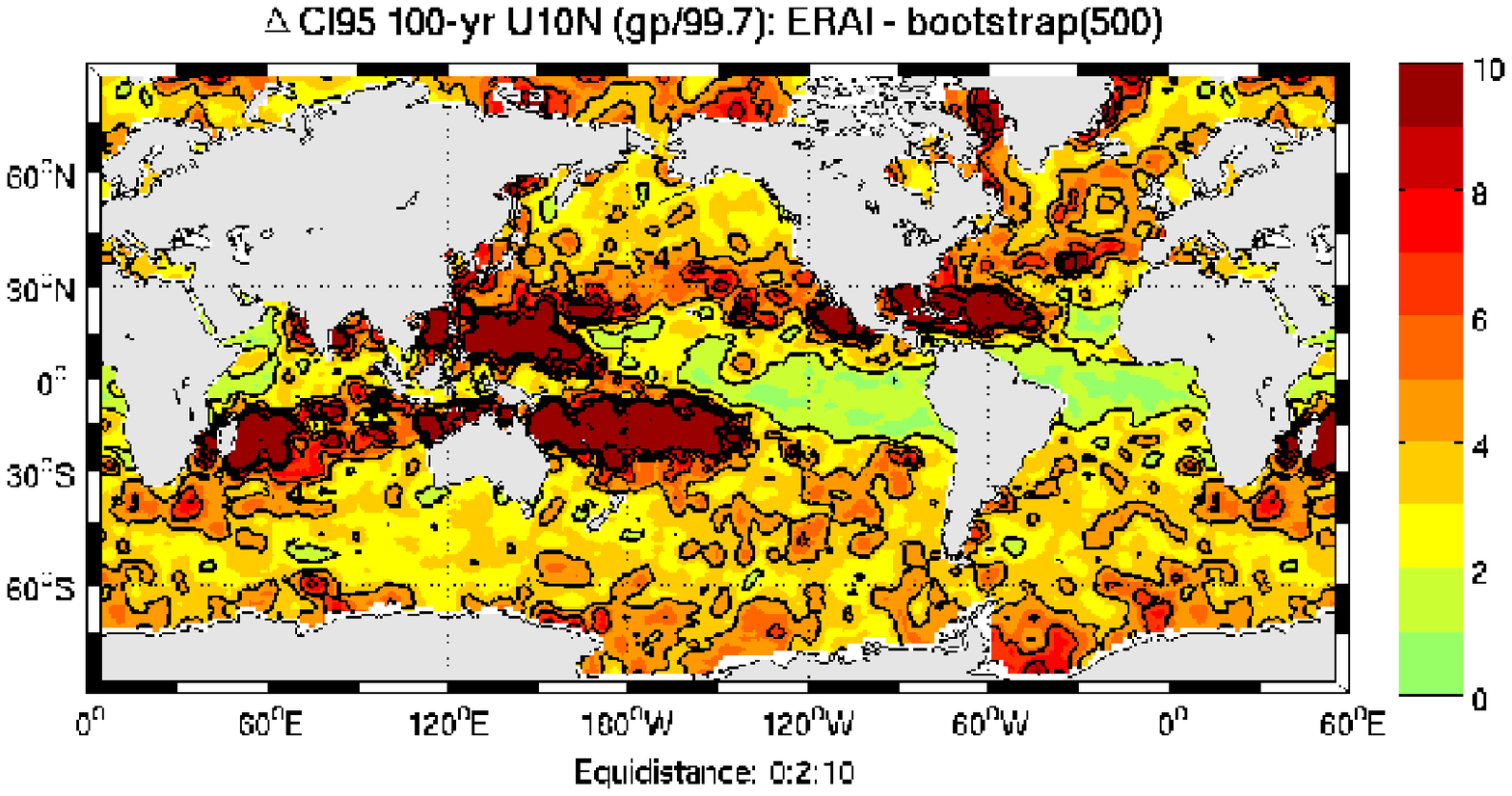}
\caption{
Panel a: The difference between \Ur~ ENS240 and ERA-I GP estimates
[m s$^{-1}$].  Panel b: The width of the confidence intervals for ERA-I GP
estimates. 
Outside the tropics the EXP and GP estimates (see \Fig{U100})
both agree well with DRE, but ERA-I GP has much wider confidence intervals
in regions with tropical cyclones.}
\label{fig:erai_gp} 
\end{center} 
\end{figure}

\begin{figure}[h]
\begin{center}
(a)\includegraphics[scale=0.7]{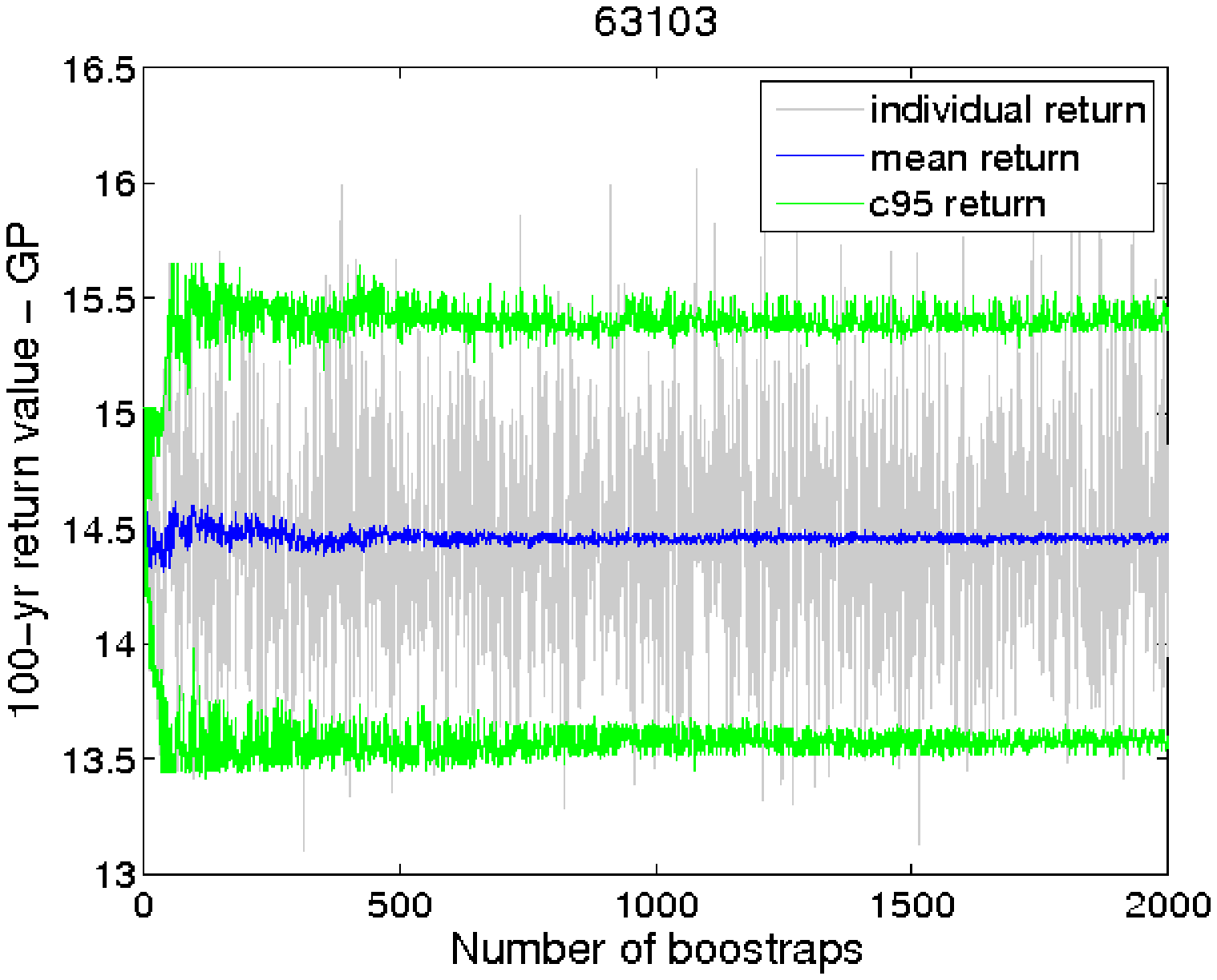}\\
(b)\includegraphics[scale=0.7]{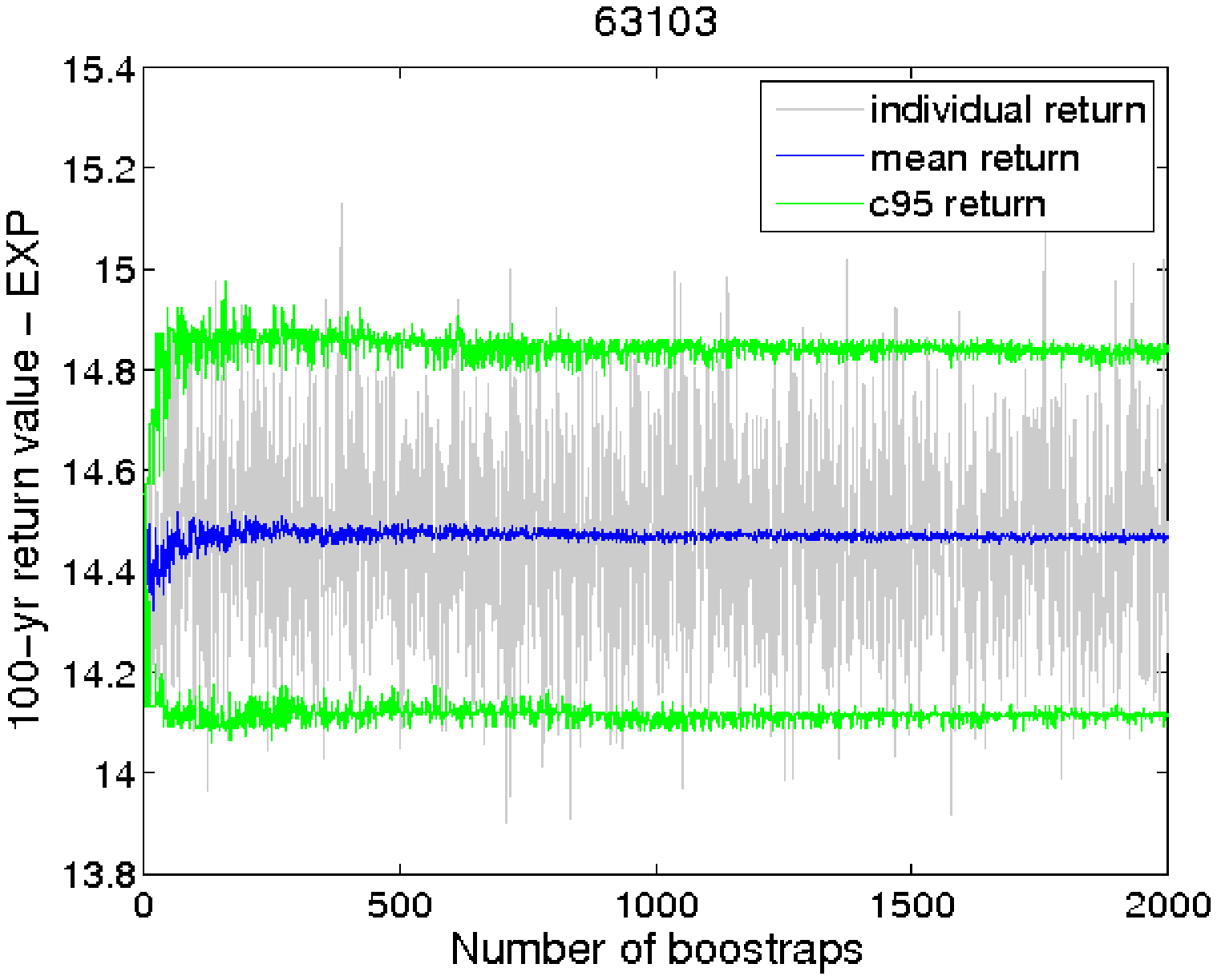}
\caption{
Panel a: Bootstrap GP confidence levels and mean return value estimates
for \Hr~ as a function of bootstrap sample size for buoy 63103 on the North
Cormorant location NW of the Shetland Isles in the southern Norwegian Sea
(61.2$^\circ$ N, 001.1$^\circ$ E). See Fig S2a for \emph{in situ} locations.
Shown in green are the 2.5\% and 97.5\% confidence levels. The bootstrap mean
return value is shown in blue. The individual estimates are shown in gray.
Panel b: Same as Panel (a) for the EXP distribution.}
\label{fig:bootsize} 
\end{center} 
\end{figure}

\begin{figure}[h]
\begin{center}
\includegraphics[scale=0.7]{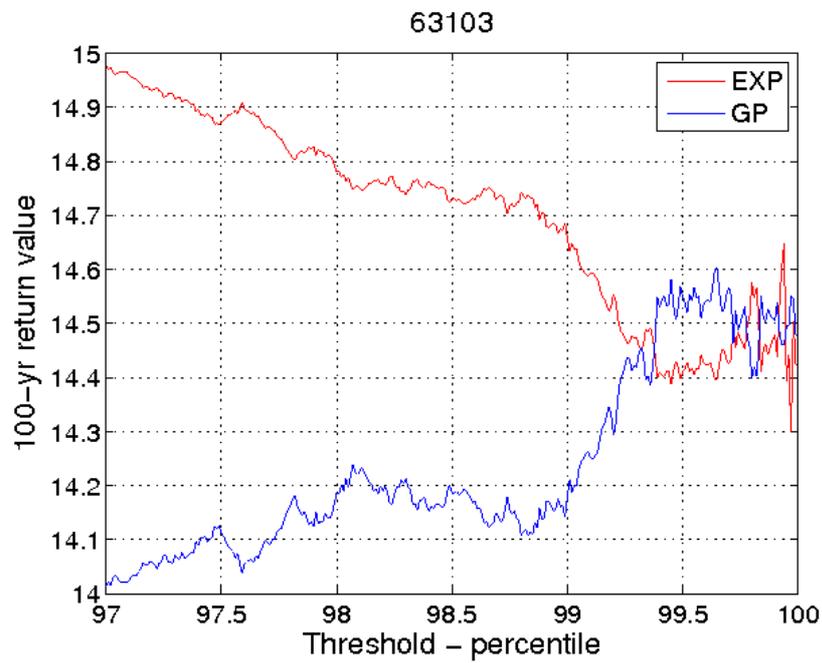}
\caption{
Return value estimates of \Hr~ as a function of threshold level for buoy
63103 in the southern Norwegian Sea.  EXP (red) and GP (blue) estimates
start to converge after 99\% and exhibit rather stable behavior around 99.7\%.}
\label{fig:threshold} 
\end{center} 
\end{figure}

\begin{figure}[h]
\begin{center}
(a)\includegraphics[scale=0.7]{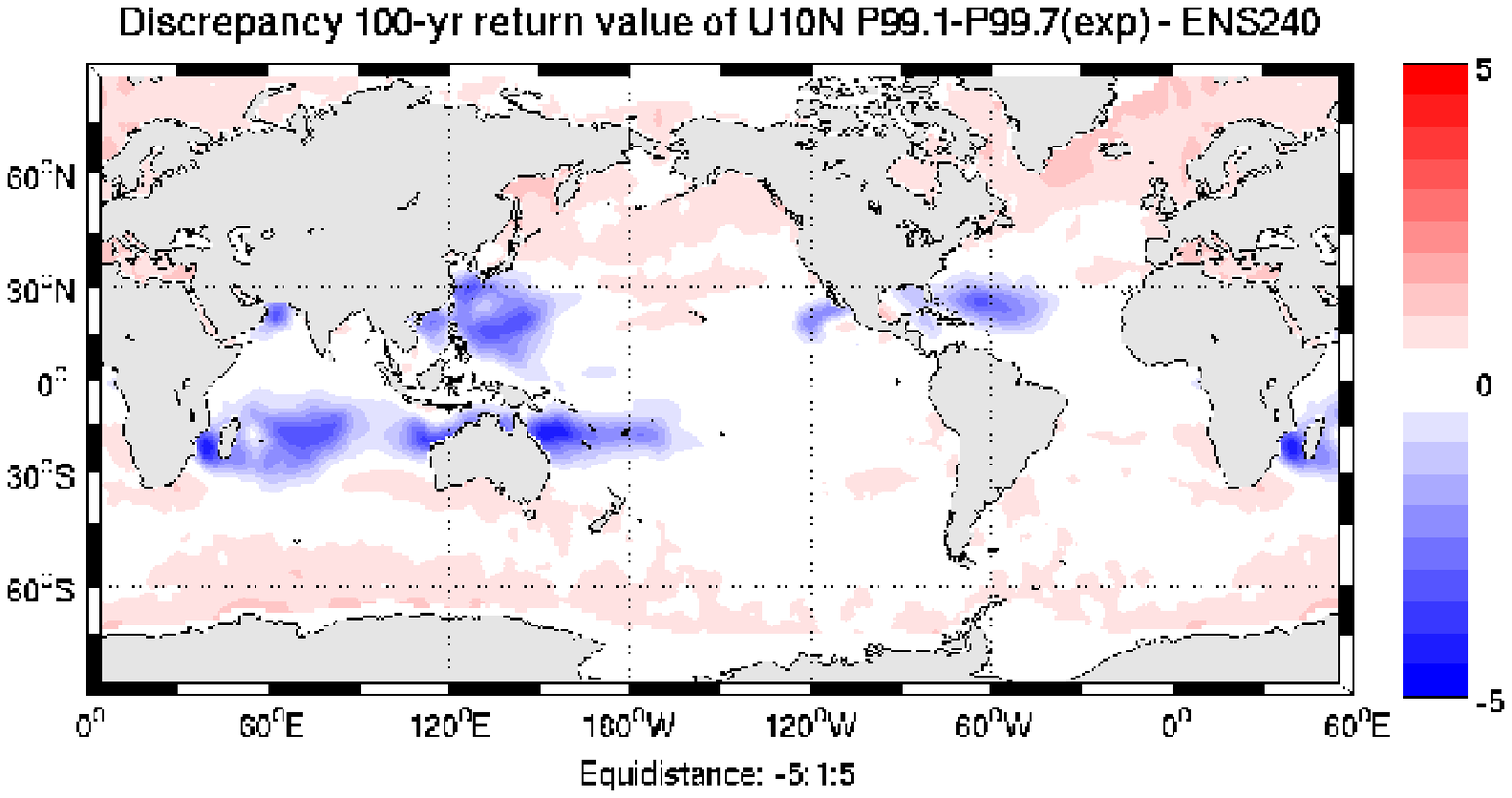}\\
(b)\includegraphics[scale=0.7]{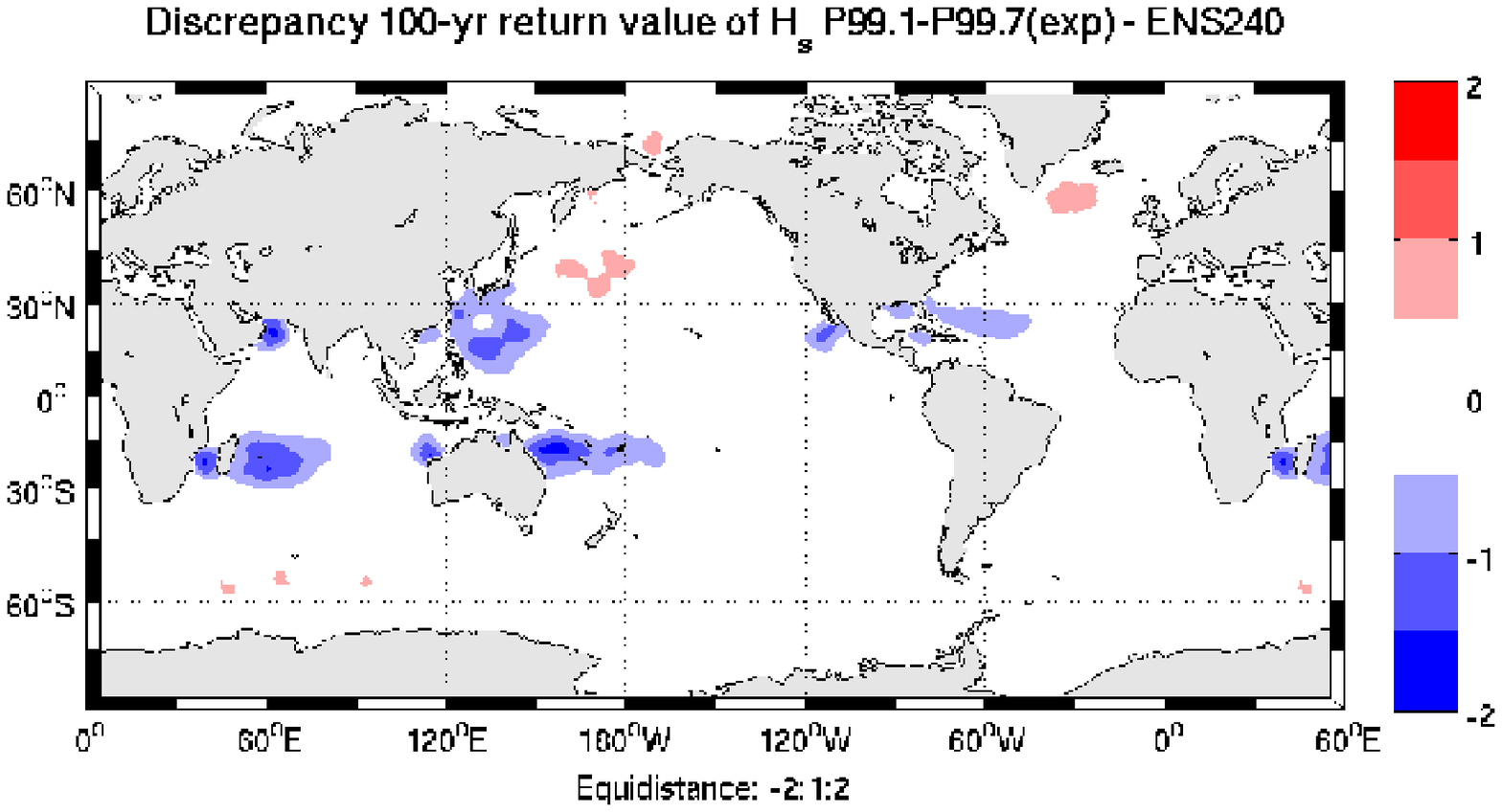}
\caption{
Panel a: Difference plot of the \Uexp~ return value estimates at
$P_{99.7}$ and $P_{99.1}$. Outside the tropics the differences are rather
small, and by investigation of the marginal (pointwise) behavior (see previous
Figure) it is clear that $P_{99.7}$ is a reasonable choice. In areas visited by
tropical cyclones the estimates show much larger deviation. Panel b: Same as
panel (a) but for \Hexp.}
\label{fig:threshmap} 
\end{center} 
\end{figure}    

\begin{figure}[h]
\begin{center}
(a)\includegraphics[scale=0.7]{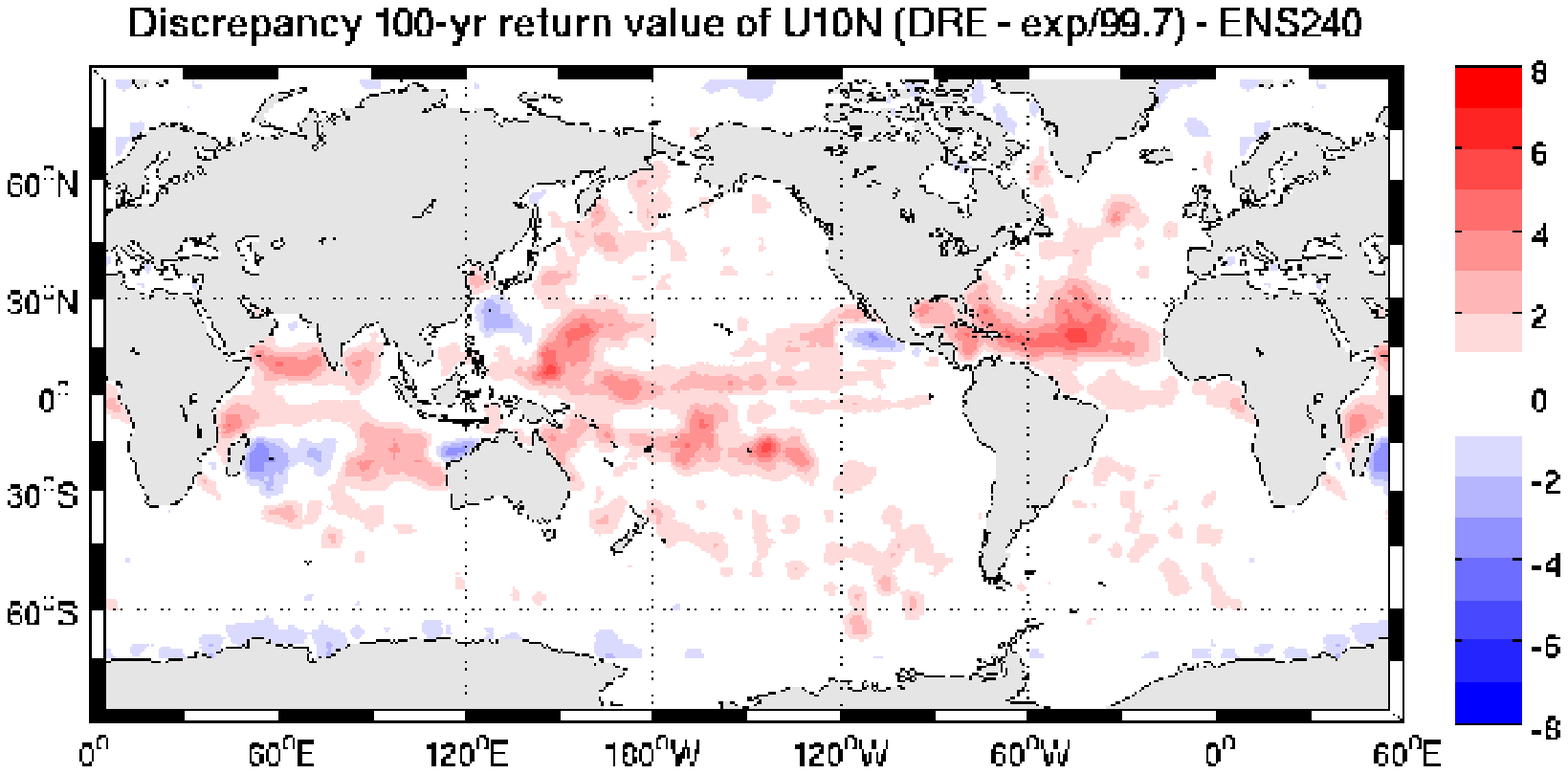}\\
(b)\includegraphics[scale=0.7]{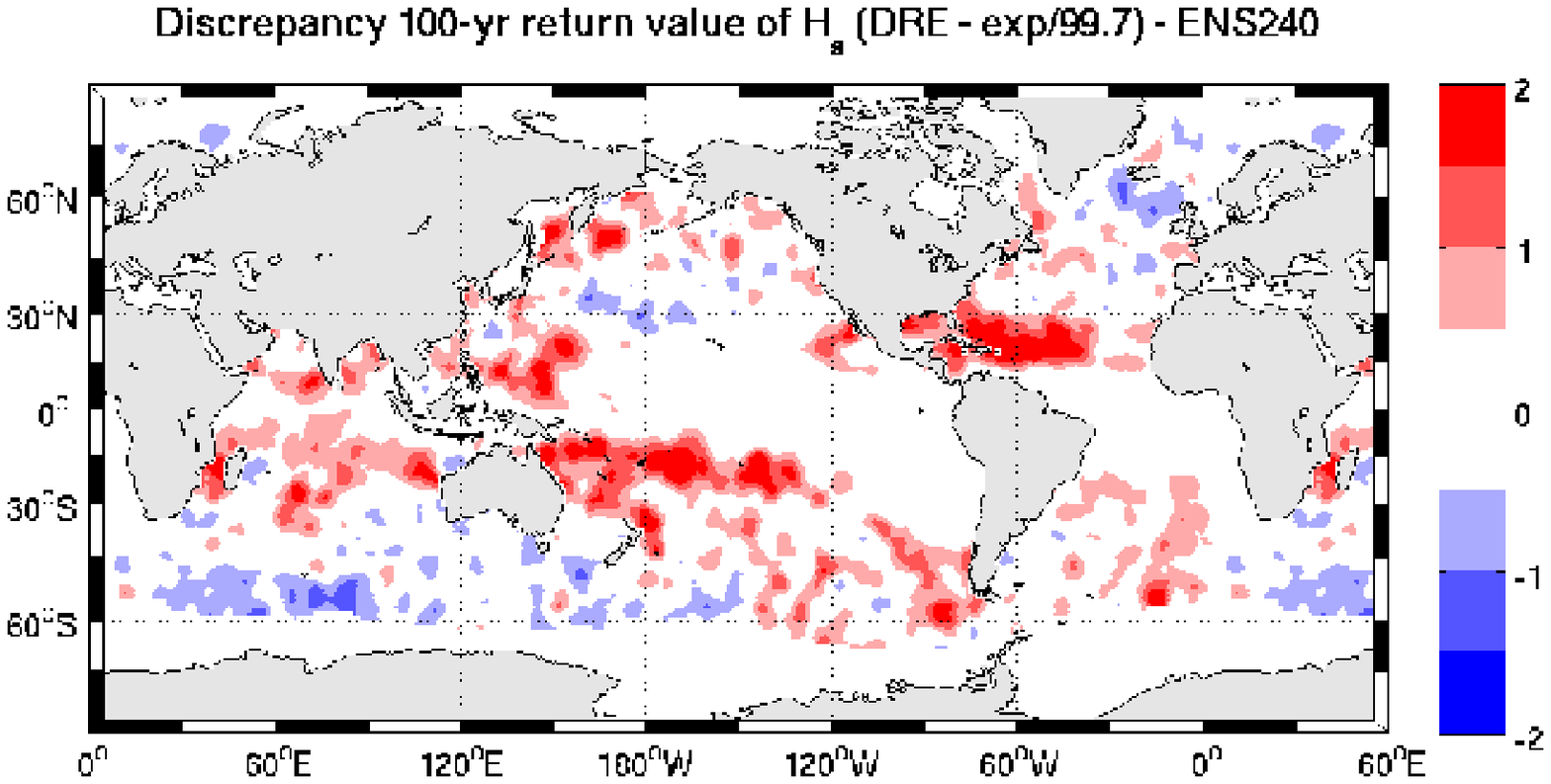}
\caption{
Panel a: The difference between DRE and the exponential (EXP) \Ur~ estimate
for ENS240 [m s$^{-1}$].  Panel b: Same as Panel a for the \Hr~ estimate [m].
Outside the tropics the EXP and GP estimates (cf \Fig{U100})
both agree well with DRE, but GP yields substantially higher return values
in regions with tropical cyclones whereas EXP does the opposite.}
\label{fig:dre_v_exp} 
\end{center} 
\end{figure}

\end{document}